\documentclass[reprint,amsmath,amssymb,pra,superscriptaddress]{revtex4-2}
\pdfoutput=1

\usepackage{amsmath}
\usepackage{bm}
\usepackage{amsfonts}
\usepackage{amssymb}
\usepackage{amsthm}
\usepackage{amstext}
\usepackage{amsbsy}
\usepackage{mathbbol}
\usepackage{dcolumn}
\usepackage{amsopn}
\usepackage{braket}
\usepackage{amscd}
\usepackage{amsxtra}
\usepackage{graphicx}
\usepackage[utf8]{inputenc}
 
\usepackage{hyperref}
\usepackage{subcaption}
\usepackage{makecell}
\usepackage{physics}

\begin{document}

\title{Two qubits in one transmon - QEC without ancilla hardware}

\author{Alexander Simm}
    \email{a.simm@fz-juelich.de}
    \affiliation{Peter Grünberg Institut - Quantum Computing Analytics (PGI-12)\\
                 Forschungszentrum Jülich, 52425 Jülich, Germany}
\author{Shai Machnes}
    \address{Qruise GmbH, 66113 Saarbrücken, Germany}
    \affiliation{Peter Grünberg Institut - Quantum Computing Analytics (PGI-12)\\
                 Forschungszentrum Jülich, 52425 Jülich, Germany}
\author{Frank K. Wilhelm}
    \affiliation{Peter Grünberg Institut - Quantum Computing Analytics (PGI-12)\\
                 Forschungszentrum Jülich, 52425 Jülich, Germany}

\date{\today}

\begin{abstract}
We show that it is theoretically possible to use higher energy levels for storing and operating two qubits within a superconducting transmon device, by  identifying its energy levels as product states between multiple effecitve qubits. As a proof of concept we show how to realise a complete set of gates necessary for universal computing by numerically optimising control pulses for single qubit gates on each of the qubits, entangling gates between the two qubits in one transmon, and an entangling gate between two qubits embedded in different transmons that are coupled. With these control pulses it is in principle possible to double the number of available qubits without any overhead in hardware. The additional qubits could be used in algorithms which need many short-living qubits such as syndrome qubits in error correction or by embedding effective higher connectivity in qubit networks. Numerical parameters are chosen to be well reachable in experiments.
\end{abstract}
\maketitle

\section{Introduction}
We are currently in the NISQ era of quantum computing in which different platforms are under active development and processors with over 50 qubits are being realised \cite{Google_2019,USTC_2021}. Further progress needs both an increase in the number of accessible qubits an an increase in fidelity of operations on these qubits. While the fidelity of the qubits under gate operations, especially two-qubit entangling gates is the primary bottleneck, increasing the number of qubits also presents challenges including overhead in control and readout hardware. A conventional approach for realising qubits is using a quantum subsystem with several energy levels and restricting the computational subspace to two of them \cite{WuLidar02}. All other states are discarded and transitions to them are considered leakage that needs to be avoided. In a superconducting device with Josephson junctions the nonlinear potential allows to address individual transitions between states which effectively separates the ground and first excited state from higher excited states. In addition to cooling and decoupling the system from the environment, control pulses are engineered to prevent leakage out of this computational subspace.

In contrast to discarding all except two states of the system, there are several ideas how to exploit them. First, higher excited states can be used as transient states for gate operations (for example the CZ \cite{Strauch03} bSWAP gates \cite{Poletto_2012} or multi-qubit gates \cite{Gu_2021}). In this case the computational subspace stays limited to two states. Another option is to include higher excited states into the computational subspace by redefining what the fundamental computational unit is: the qubit is replaced by an N-level qudit\cite{Wang_2020}. Qudits have been shown to be controllable \cite{Jirari_2009} and gates have been realised, including an entangling gate between two qudits up to dimension 5 \cite{Hrmo_2022}. A disadvantage of this approach is that known algorithms might not be easily benefit from qudits. Another application is to use nonlinear resonators as continuous variables, i.e., fundamentally change the encoding scheme \cite{Weedbrook12,Braunstein05,Hillmann20,Campagne20}. 

We propose a scheme to stay with the concept of qubits but to use more than two energy levels for storing more than one qubit in a superconducting device. This is simply done by relabeling the energy levels as product states, i.e., interpreting a subspace of dimension $2^n$ and usually composed as a direct sum as a direct product of Hilbert spaces. As a proof of concept we use four levels in a model for superconducting transmons (ground state and three excited states) to store two qubits. This concept makes sense, if the resulting effective qubit lattice has the appropriate connectivity, i.e., if we can make gates between different qubit realised in distinct transmons. This scheme has previously been proposed in \cite{Kiktenko_2015} where it was shown how iSWAP and Hadamard gates can be realised by decomposing them into direct transitions between the four energy levels. We show that it is possible to realise a complete set of gates necessary for universal computing by numerically optimising control pulses for single qubit gates on each of the qubits, entangling gates between the two qubits within one transmon, and an entangling gate between two qubits from two coupled transmons.
With these control pulses it is in principle possible to double the number of available qubits without any overhead in hardware, although the signal generation is more complicated than in usual two-level systems, as will be explained below. The additional qubits could be used in algorithms which need many qubits with large local connectivity, such as qubits in error correction. This is not as extreme as molecular quantum computing as proposed in Ref. [\onlinecite{Tesch02}] which aims at putting many qubits into a single degree of freedom. This paper complements the work \cite{Fischer_2022} as it is giving explicit pulse constructions. 

The paper is organized as follows: In section \ref{section:model} we review the model of the transmon and the numerical methods we used for finding optimal pulse shapes. Section \ref{section:single_qubit_gates} and \ref{section:single_transmon_entangling_gate} explain the results for single- and two-qubit gates within a transmon. Entangling gates between two and four qubits in two coupled transmons as well as the method we used to find those gates are explained in section \ref{section:two_transmons}. All parameter values and plots of the propagators can be found in appendix \ref{section:appendix_parameters}.
\section{Model}
\label{section:model}
\subsection{Transmon and drive Hamiltonian}
\label{subsec:model_hamiltonians}
We consider a superconducting device containing a single Josephson junction that is described by the Hamiltonian \cite{Krantz_2019}
\begin{align}
    H_0 = 4 E_c n^2 - E_J \cos(\phi)
    \label{eq:drift_hamiltonian_superconductor}
\end{align}
where $n$ is the number of Cooper pairs on the effective capacitance and $E_c$ is the charge energy required to add an electron there. The Josephson junction contributes the cosine potential with energy $E_J$ and the flux $\phi$. In the transmon limit, $E_J \gg E_c$, the cosine can be expanded around its minimum at $\phi=0$. Up to the quartic term this leads to an anharmonic oscillator \cite{Krantz_2019}
\begin{align}
    H_0 = \omega a^{\dagger} a + \frac{\lambda}{2}a^{\dagger} a^{\dagger} a a
    \label{eq:drift_hamiltonion_oscillator}
\end{align}
where $a,a^{\dagger}$ are the bosonic operators for the Cooper pairs, $\omega = \sqrt{8 E_c E_J} - E_c$ is the resonance frequency between ground and first excited state, and $\lambda = -E_c = E_2 - 2E_1 < 0$ denotes the anharmonicity. In this notation the approximation is valid in the limit $\lambda \ll \omega$, which is justified for transmons that operate in the range of $\omega \sim 3\dots 6$ GHz and $|\lambda| \sim 100 \dots 300$ MHz. In the following, this anharmonic oscillator will simply be called the transmon \cite{Koch_2007}.

In addition to the drift $H_0$, the full system $H(t) = H_0 + H_d(t)$ contains the time-dependent drive
\begin{align}
    H_d = A v(t) (a + a^{\dagger})
    \label{eq:drive_hamiltonion}
\end{align}
which models the microwaves used to control superconducting qubits. It consists of a constant amplitude scale factor $A$ and a dimensionless time-dependent function
\begin{align}
    v(t) = s(t) \cos(\omega_d t + \phi_d) = s(t) \left( I \cos(\omega_d t) - Q \sin(\omega_d t) \right)
    \label{eq:drive_function}
\end{align}
with constant in-phase component $I = \cos(\phi_d)$ and quadrature component $Q = \sin(\phi_d)$. The system is thus driven by a carrier signal at a frequency $\omega_d$ with additional phase shift $\phi_d$. The signal is shaped by an envelope $s(t)$ which will be the main object of optimisation. Typical intuitive choices for the envelope are Gaussians or piecewise constant functions.

%{\bf are $I$ and $Q$ constant? If so, why?}
%{\\\bf\textit{For some reason the C3 code was written to assume the envelope $s(t)$ from the AWG to be real. I am running the simulations again with $s$ complex, i.e. $v(t) = \Re(s(t)) \cos(\omega_d t + \phi_d) + \Im(s(t)) \sin(\omega_d t + \phi_d)$.\\}}
%we should get back to this 

On occasion, the DRAG scheme is used to eliminate unwanted transitions out of the computational subspace or between qubit subspaces \cite{Motzoi_2009}. It transforms the envelope function into $s(t) \to s(t) - i\delta \frac{\dot s(t)}{\lambda}$ where $\delta \ll 1$ is a free parameter that scales the correction term.

\begin{figure}[t]
        \centering
        \includegraphics[width=0.65\linewidth]{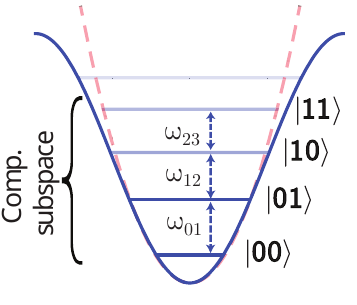}
        \caption{Labeling of the energy levels in the computational subspace (modified from \cite{Krantz_2019})}
        \label{fig:model_computational_subspace}
\end{figure}

In order to store two qubits we use the four lowest energy levels (Fig. \ref{fig:model_computational_subspace}) identified as the product states of two qubits. We will denote the qubits as $Q_1$ ($Q_2$), referring to the left (right) qubit in the ket $\ket{Q_1 Q_2}$. States in this relabeled 2-qubit basis will be written as normal kets, while states in the transmon's eigenbasis will be written with a bar. The mapping is therefore $\ket{\Bar{0}} = \ket{00}$, $\ket{\Bar{1}} = \ket{01}$, $\ket{\Bar{2}} = \ket{10}$, and $\ket{\Bar{3}} = \ket{11}$. Higher excited states $\ket{\Bar{n}}$, $n>3$, are considered leakage and will only be denoted by the eigenstate ket. In the simulations one additional fifth level $\ket{\Bar{4}}$ was included to simulate leakage out of the computational subspace. The energy eigenstates $\ket{\Bar{n}}$ have eigenvalues $E_n = n\omega - \frac{n(n-1)}{2}\lambda$. The Bohr frequency between two levels will be denotes as $\omega_{m\to n} = |E_n - E_m|$.

This labelling is compatible with the usual computational subspace in the lowest two levels. As long as the additional qubit is not needed, algorithms can fall back to using two levels, effectively assuming that $Q_1$ is in state $\ket{0}$.

Projected onto the computational subspace, the system can be written in the basis $\{\mathbb{1}, \sigma_x, \sigma_y, \sigma_z\}^{\otimes 2}$ of $SU(4)$, i.e. in tensor products of Pauli matrices and the unit matrix. The Hamiltonian, eq.  (\ref{eq:drift_hamiltonion_oscillator}), then becomes
\begin{equation}
\begin{aligned}
  H_0 &= \left(\frac{3}{2}\omega + \lambda\right) \mathbb{1}\otimes\mathbb{1} ~-~ \frac{1}{2}( \omega+\lambda) \mathbb{1}\otimes Z \\&-~ (\omega+\lambda) Z\otimes\mathbb{1} ~+~ \frac{\lambda}{2} Z\otimes Z
  \label{eq:drift_hamiltonian_paulis}
\end{aligned}
\end{equation}
while the drive, eq. (\ref{eq:drive_hamiltonion}), is
\begin{equation}
\begin{aligned}
  H_d &= A v(t) \Big[ \frac{1}{2}(1+\sqrt{3}) \mathbb{1}\otimes X ~+~ \frac{1}{2}(1-\sqrt{3}) Z\otimes X \\&+~ \frac{1}{\sqrt{2}}(X\otimes X + Y\otimes Y) \Big]
  \label{eq:single_transmon_drive}    
\end{aligned}
\end{equation}

This structure does not lend itself to an easy and straightforward adaptation of simple and intuitive control strategies. We thus rather resort to optimal control \cite{Roadmap_2015,Roadmap_2022}.

\subsection{Numerical optimisation}
\label{subsec:numerical_optimisation}
In a finite system, calculating the propagator $U(T) = \mathcal{T} \exp\left(-i\int_0^T H(t) dt\right)$ for a given drive Hamiltonian $H_d$ and a gate time $T$ is usually only feasible numerically. Assuming the Hamiltonian is (or can be approximated as being) piece-wise constant, this is a matter of matrix exponentiation and multiplication along the time ordering described by $\mathbb{T}$. We are attempting to solve the inverse problem of finding a drive Hamiltonian $H_d$ that generates a given gate. This is done with numerical optimisation: start with an initial guess for $H_d$ that contains enough parameters, calculate $U$, and optimise the parameters with respect to a goal function $g(U, G)$ that calculates the distance between $U$ and the ideal propagator (i.e. the desired gate) $G$. If not otherwise mentioned, the goal function $g = 1 - F$ was used with the fidelity $F = \frac{1}{d} tr(U^{\dagger} G)$ where $d$ is the system's dimension. The fidelity is only evaluated in the computational subspace with $d=4$, which in particular means that relative phases in higher excited states are being ignored. If the optimisation reaches a global (or local, but good enough) minimum of $g$, the drive Hamiltonian realises the ideal gate $G$ up to a small error. For all following results the numerical optimisation was done using the $C^3$ software \cite{Wittler_2021} with gradient-based optimisation (L-BFGS). All simulations are done assuming a closed system at zero temperature without noise.

In addition to the optimisation algorithm, the $C^3$ toolkit simulates the chain of devices which would create the drive function $s(t)$ in a lab. The chain of devices includes a local oscillator (LO) that generates the carrier signal $\cos(\omega_d t + \phi_d)$. In the following, $\omega_d$ will be chosen to be resonant to a desired transition between two energy eigenstates. Additionally, a waveform generator (AWG) generates the envelope $s(t)$, which is mixed with the carrier signal into the drive function (\ref{eq:drive_function}). 

The set of optimisable parameters thus contains the scaling amplitude of the drive $A$, the gate time $T$, the drive frequency $\omega_d$ and a phase shift $\phi_d$, the strength of the DRAG correction $\delta$, and all parameters that specify the envelope $s(t)$.

%Since a gate time of time $T$ correponds to a spectral resolution of roughly $\frac{1}{T}$, any time $T \lambda \gg 1$ (i.e. $T \gg 3.3$ ns)
%{\bf keep dimensionless as long as possible.}
%{\\\bf \textit{$T \lambda \gg 1$ instead of $T \gg \frac{1}{\lambda}$ ?}\\} 
%should suffice for driving a specific transition without affecting other first-order resonances. %However, as this only serves as a proof of concept, we did not try to find optimal gate times and %used comparably long gate times to avoid any unwanted transitions.

In order to compare the optimised drive frequencies with the resonances of the model we numerically calculated the Stark shifted eigenvalues of the full Hamiltonian (\ref{eq:hamiltonian_two_transmons}).
Since the drive amplitude is not constant, the Stark shift changes over time. We therefore used the Hamiltonian at $\frac{T}{2}$ which, in case of the Gaussian envelopes, contains the strongest shift.
Although this procedure only approximates the actual Stark shift, it is sufficient for comparing the optimised frequencies to the resonances, especially because the Stark shift is only in the order of a few MHz.

\subsection{Alignment of phases}
\label{subsec:phase_alignment}
Under time evolution with $U_0 = e^{-i H_0 t}$ the free system picks up dynamical phases which need to be corrected for the gates. In contrast to the case of two level systems, there is no transformation into a rotating frame that removes both the drift and the oscillation of the drive. A transformation $T = e^{-i H_0 t}$ removes the drift, but creates higher order terms in the drive, because the drive does not commute with the anharmonicity. Instead we can do the transformation $T = e^{-i\omega t a^{\dagger} a}$ with the harmonic part of the drift \cite{Schneider_2018}. This results in
\begin{align}
    \tilde H &= T^{\dagger} H T + i T^{\dagger} \dot T \\
     &= \frac{\lambda}{2} a^{\dagger} a^{\dagger} a a ~+~ A s(t) (a + a^{\dagger})
\end{align}
where the oscillation of the drive disappeares after a rotating wave approximation (compare eq. (\ref{eq:drive_function})) but the anharmonicity remains. The second and third excited state still pick up phases under time evolution. We thus have the option of actively correcting phases with the drive Hamiltonian or of choosing gate times at which the phases of the computational subspace align. The latter could be achieved by allowing the gate time $T$ to be numerically optimised which leads to times at which at least some of the phases vanish. In order to have all phases of the excited states aligned with the ground state we need to find $n,m\in \mathbb{Z}$ such that in the rotating frame (frequencies are in units of $\frac{2\pi}{s}$)
\begin{equation}
\begin{aligned}
    E_2 T &= \lambda T = n \\
    E_3 T &= 3\lambda T = 3\lambda m \\
    E_3 T &= 3(\omega-\lambda) T = l
\end{aligned}
\end{equation}
The first condition fixes the time to $T = \frac{n}{\omega}$, while the last two lead to $m = (2 - \frac{\lambda}{\omega})n$ and $l = 3(1 - \frac{\lambda}{\omega})n$. Unless $n$ is large enough so that $\frac{\lambda}{\omega}n$ is integer, the phases will in general only partially align and need to be actively corrected. For the chosen values of $\omega=5$ GHz, $\lambda=300$ MHz, however, $n,m,l$ are integer for $T$ being a multiple of 10 ns. If this seems arbitrary, the argument works the other way around as well: using a tunable transmon it would be possible to fix $T$ and optimise its anharmonicity until the phases (mostly) align.
%{\bf is there a way to AC-Stark-shift the phases away?}
%{\\\bf\textit{I think it is already included in the optimised results. I added the following paragraph. Is it sufficient as an explanation?\\}}
Perfect alignment of the drift phases is lifted by the AC-Stark shift caused by the drive. For some gates the shift is in the range of several MHz which, depending on the gate time, can correspond to almost a full rotation. However, since the amplitude and the carrier frequencies are variable parameters for the optimisation we expect the results to already be adjusted to the shifted energies.

\begin{figure*}[t]
    \centering
    \begin{subfigure}[b]{0.45\textwidth}
        \centering
        \includegraphics[width=\textwidth]{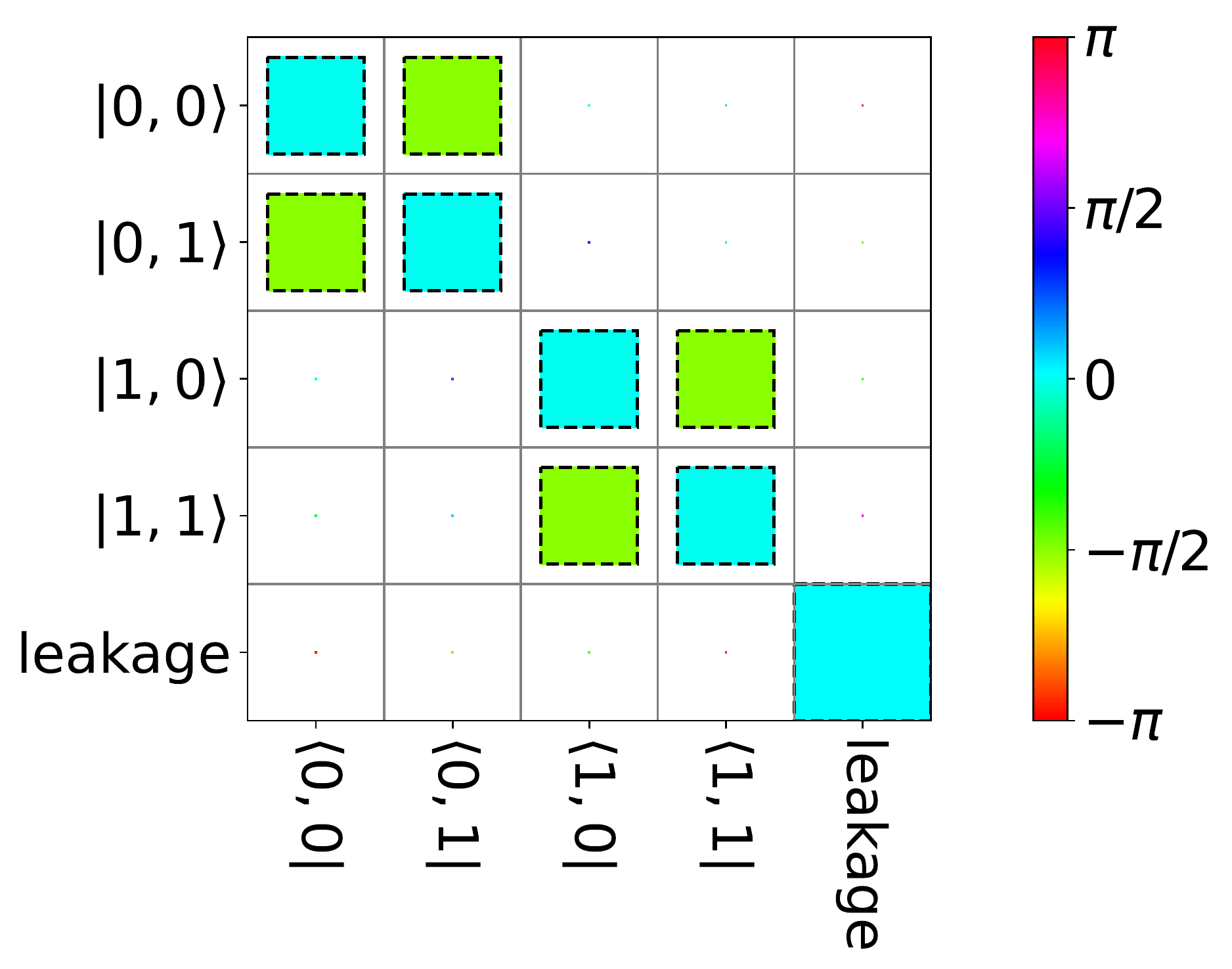}
    \end{subfigure}
    \hfill
    \begin{subfigure}[b]{0.45\textwidth}
         \centering
         \includegraphics[width=\textwidth]{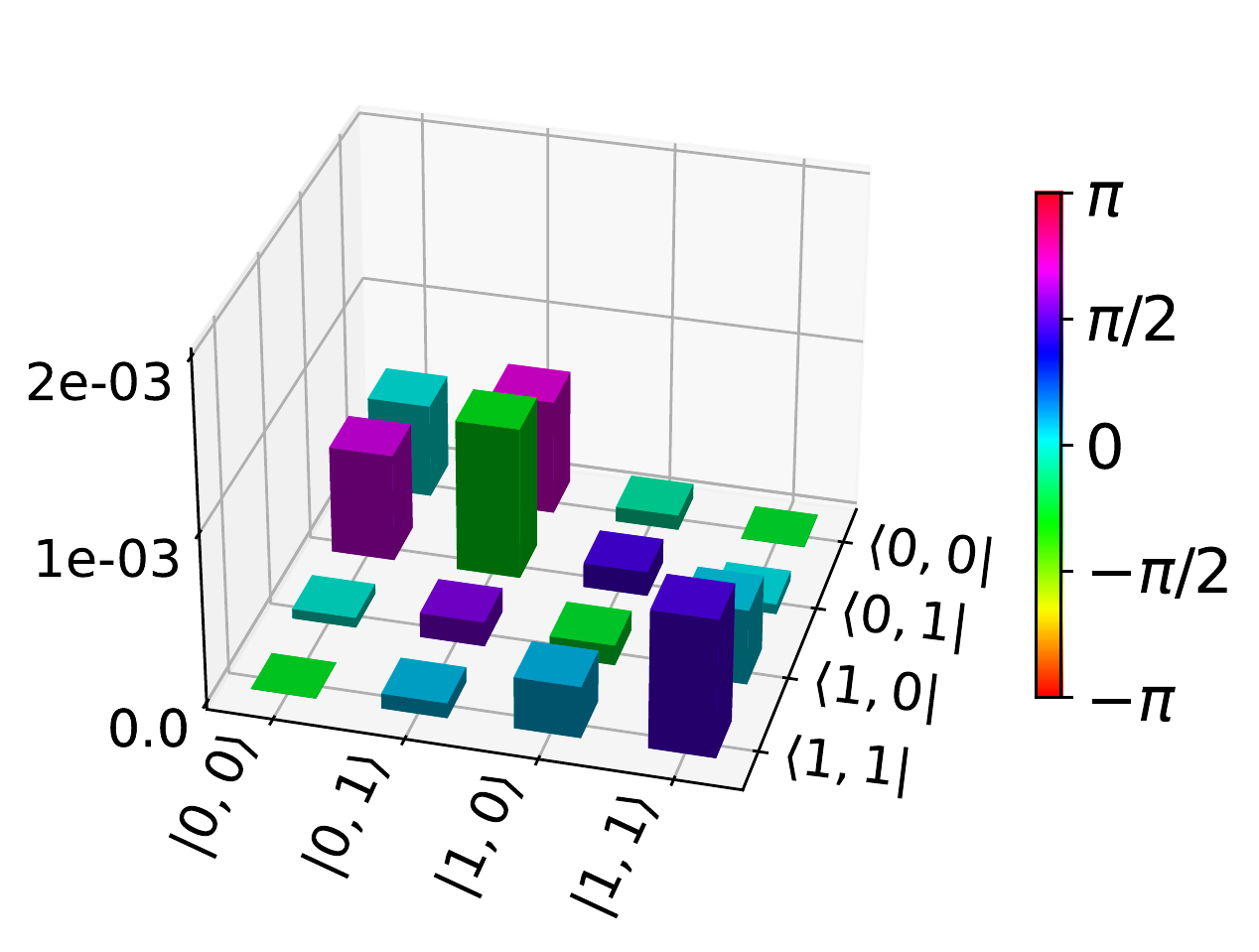}
    \end{subfigure}
    \\
    \begin{subfigure}[b]{0.48\textwidth}
        \includegraphics[width=\textwidth]{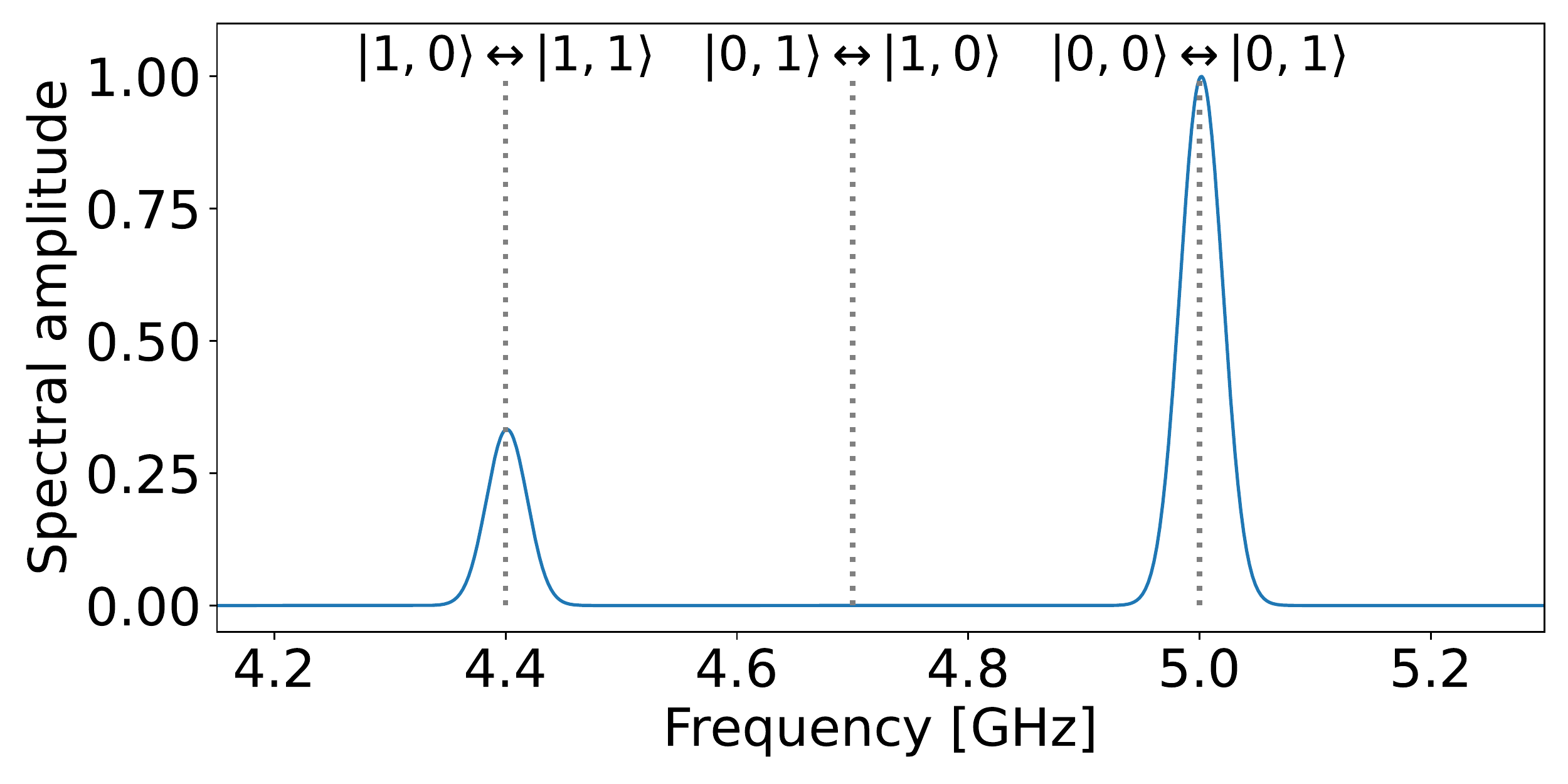}
    \end{subfigure}
    \caption{$X(\pi/2)$ gate on qubit $Q_2$. Colours indicate the phase and areas are proportional to the absolute amplitude. \textbf{(a)} The optimised propagator $U$. Dotted black squares indicate the absolute values of the ideal gate. \textbf{(b)} the difference $Ue^{i\phi} - G$ between $U$ and the ideal gate $G$. The phase difference $\phi = \arctan(U_{0,0}) - \arctan(G_{0,0})$ corrects the global phase which is ignored in the optimisation. \textbf{(c)} Normalised spectral amplitude of the optimised pulse. Dotted lines correspond to resonances of the model.}
    \label{fig:single_transmon_rx90p_q2}
\end{figure*}

\section{Single qubit gates}
\label{section:single_qubit_gates}

The simplest gates that we wish to realise are single-qubit gates on each of the qubits within one transmon, which corresponds to splitting up $SU(4)$ into $SU(2)\otimes SU(2)$. For the simulation of all of these gates, we used parameters $\omega=5$ GHz and $\lambda=300$ MHz which are in the typical range for transmons. As described in section \ref{subsec:phase_alignment} these values force us to use gate times $T$ in multiples of $10$ ns unless we are actively correcting the phases. For all single qubit gates, the envelopes are unnormalised Gaussians
\begin{align}
    s(t) = \exp\left(-\frac{(t - t_0)^2}{2 \sigma^2}\right)
\end{align}
that contain $t_0$ and $\sigma$ as optimisable parameters. This means that they are not considered to be piece-wise constant but continuous (up to the simulation's resolution of 0.02 ns). Initial values for the optimisation are $t_0 = \frac{T}{2}$ and $\sigma = \frac{T}{5}$. The gates do not depend on the actual shape of the envelope, but the action $\int_0^T H_d(\tau) d\tau$ depends on the integrated area under the envelopes. Gates could thus be made shorter by using rectangular or flat-top Gaussian shapes, which we did not attempt here. All optimised parameter values are listed in Appendix \ref{section:appendix_parameters}.

\subsection{Gates on qubit \texorpdfstring{$Q_2$}{}}
\label{subsec:single_qubit_gates_Q2}

For gates on $Q_2$ the two subspaces spanned by $\{\ket{\Bar{0}}, \ket{\Bar{1}}\}$ and $\{\ket{\Bar{2}}, \ket{\Bar{3}}\}$ need to be treated separately and transitions between them have to be avoided (see fig. \ref{fig:model_computational_subspace}). Thus, two transitions $\ket{\Bar{0}} \leftrightarrow \ket{\Bar{1}}$ and $\ket{\Bar{2}} \leftrightarrow \ket{\Bar{3}}$ need to be addressed simultaneously. This was not possible with one carrier signal and a simple envelope. We chose to use a total of two drive signals. Formally, we replaced the drive Hamiltonian (\ref{eq:drive_hamiltonion}) by
\begin{align}
    H_d = \sum_{k=1,2} A_k s_k(t) \cos\left(\omega_{d}^{(k)} t + \phi_d^{(k)}\right) \,(a + a^{\dagger})
    \label{eq:drive_hamiltonian_superposition}
\end{align}
This means that there is a carrier signal and envelope for each frequency that needs to be driven, and both signals are superposed in the end, which allows controlling both transitions individually. From an experimental perspective, a single QI mixer is sufficient, with the LO set between 4.4 and 5 GHz, and the required shifts are handled by adding shifts to the envelope.

Both carrier frequencies were chosen resonant to $\omega_{0\to 1} = 5$ GHz and $\omega_{2\to 3} = 4.4$ GHz, respectively. Since phases are already corrected by the fixed gate time and sufficiently narrow peaks allow driving the chosen transitions only, the $\mathbb{1}\otimes X(\pi/2)$ gate with $X(\pi/2) = e^{i \frac{\pi}{4} \sigma_x} = \frac{1}{\sqrt{2}}(\mathbb{1} + \sigma_x)$
is the easiest to realise. After optimisation, the infidelity in the subspace is $1 - F < 10^{-6}$. The corresponding propagator is depicted in figure \ref{fig:single_transmon_rx90p_q2}. A $Y(\pi/2)$ gate on $Q_2$ can be realised the same way since the distinction between $X$- and $Y$-rotations is only a matter of shifting the phase $\phi_d$ by $\frac{\pi}{2}$.

Although on superconducting qubits $Z$-gates can be realised as virtual gates by means of shifting the phase between two gates \cite{McKay_2017}, this would have a different effect in the $SU(2)\otimes SU(2)$ basis and it was necessary to implement the gate explicitly. In this case it is only necessary to introduce relative phases on the diagonal of the propagator. We realised this by driving both transitions $\ket{0,0} \leftrightarrow \ket{0,1}$ and $\ket{1,0} \leftrightarrow \ket{1,1}$ to a full $2\pi$ rotation during which the phases were corrected by the relative phases of the two drive signals. Again an infidelity $1-F < 3\cdot 10^{-5}$ was reached.

\subsection{Gates on qubit \texorpdfstring{$Q_1$}{}}
\label{subsec:single_qubit_gates_Q1}
\begin{figure}[t]
    \includegraphics[width=0.5\textwidth]{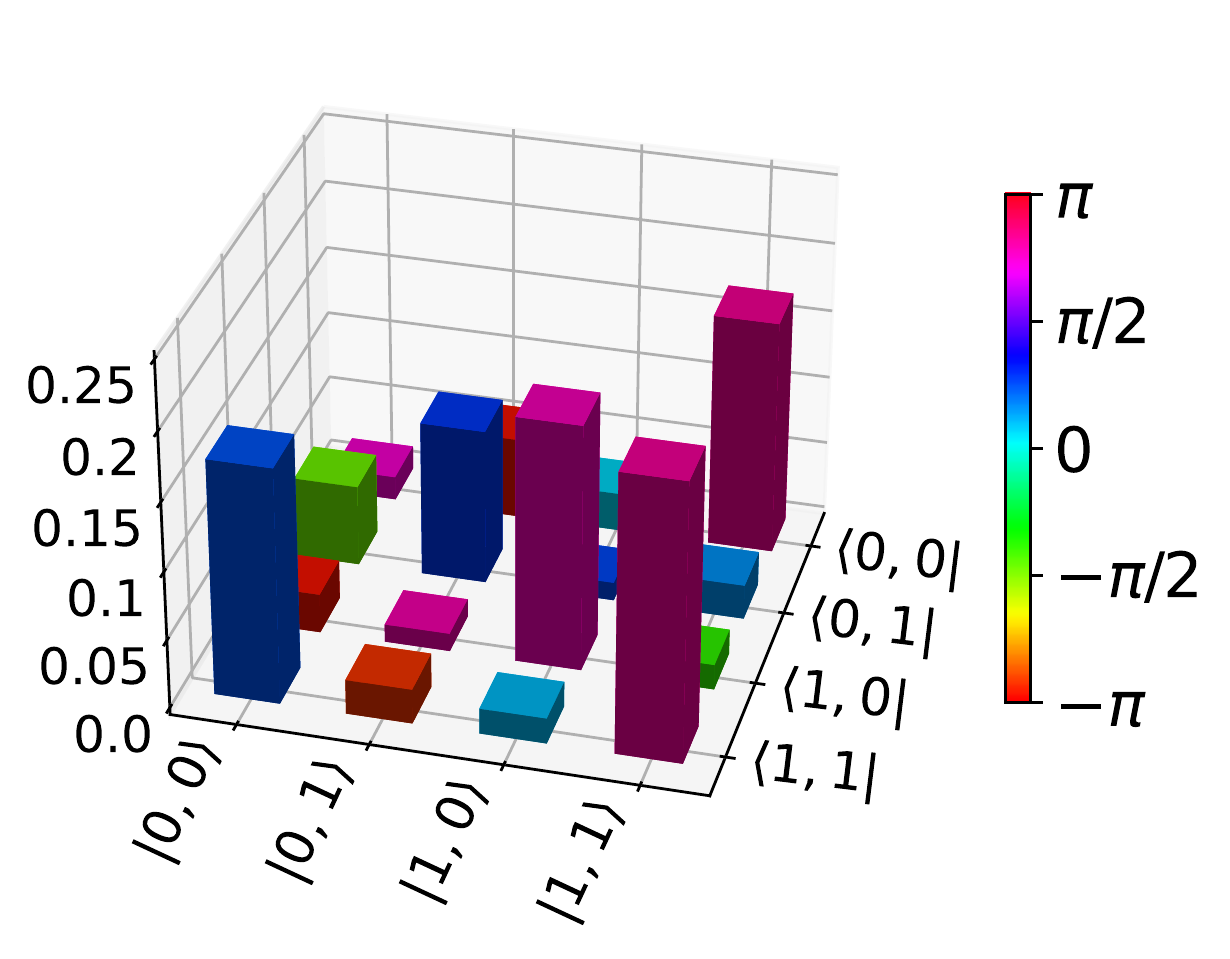}
    \caption{Error of the $Z(\pi/2)$ gate on qubit $Q_1$ computed as the difference $Ue^{i\phi} - G$ between the optimised gate $U$ and the ideal gate $G$. The phase difference $\phi = \arctan(U_{0,0}) - \arctan(G_{0,0})$ corrects the global phase which is ignored in the optimisation.}
    \label{fig:single_transmon_rz90p_q1_error}
\end{figure}

Gates on qubit $Q_1$ need transition between next-to-nearest-neighbour energy levels $\ket{\Bar{0}} \leftrightarrow \ket{\Bar{2}}$ and $\ket{\Bar{1}} \leftrightarrow \ket{\Bar{3}}$. Driving the frequencies $\omega_{0\to 2}$ and $\omega_{1\to 3}$, which are above 9 GHz, caused many unwanted 2-photon transitions which we could not remove by optimisation. Similarly, driving the three nearest-neighbour transitions $\omega_{0\to 1}$, $\omega_{1\to 2}$, and $\omega_{2\to 3}$ did not result in a high fidelity because the two intended transitions overlap in the middle (in $\ket{\Bar{1}} \leftrightarrow \ket{\Bar{2}}$). The best fidelity was achieved with two-photon transitions by driving $\frac{\omega_{0\to 2}}{2}$ and $\frac{\omega_{1\to 3}}{2}$ with a strong amplitude. Since these frequencies are at least $\frac{\lambda}{2}$ distant from all nearest-neighbour transitions, they are sufficiently off-resonant to not cause single-photon absorption. In addition to a larger amplitude the optimisation resulted in non-vanishing phases $\phi_d$. As before we only realised the $X(\pi/2)\otimes\mathbb{1}$ and $Z(\pi/2)\otimes\mathbb{1}$ gates, which resulted in errors $1-F < 2 \cdot 10^{-3}$ and $1-F \approx 0.026$, respectively. Their propagators are depicted in figures \ref{fig:appendix_single_transmon_rx90p_q1} and \ref{fig:appendix_single_transmon_rz90p_q1}.

As an example, the error of the $Z(\pi/2)$-gate, which has the lowest fidelity of all single qubit gates, can be seen in figure \ref{fig:single_transmon_rz90p_q1_error}. The main contributions are phase errors on the diagonal, probably resulting from imperfect rotations by $2\pi$ which are needed for the $Z$-gate. In addition, there are unwanted 3-photon absorptions from $\ket{00}$ to $\ket{11}$, most likely due to the large drive amplitude. The errors in all other gates are similar.

\section{Entangling gates between \texorpdfstring{$Q_1$}{} and \texorpdfstring{$Q_2$}{}}
\label{section:single_transmon_entangling_gate}
\begin{figure*}[t]
    \centering
    \begin{subfigure}[b]{0.45\textwidth}
        \centering
        \includegraphics[width=\textwidth]{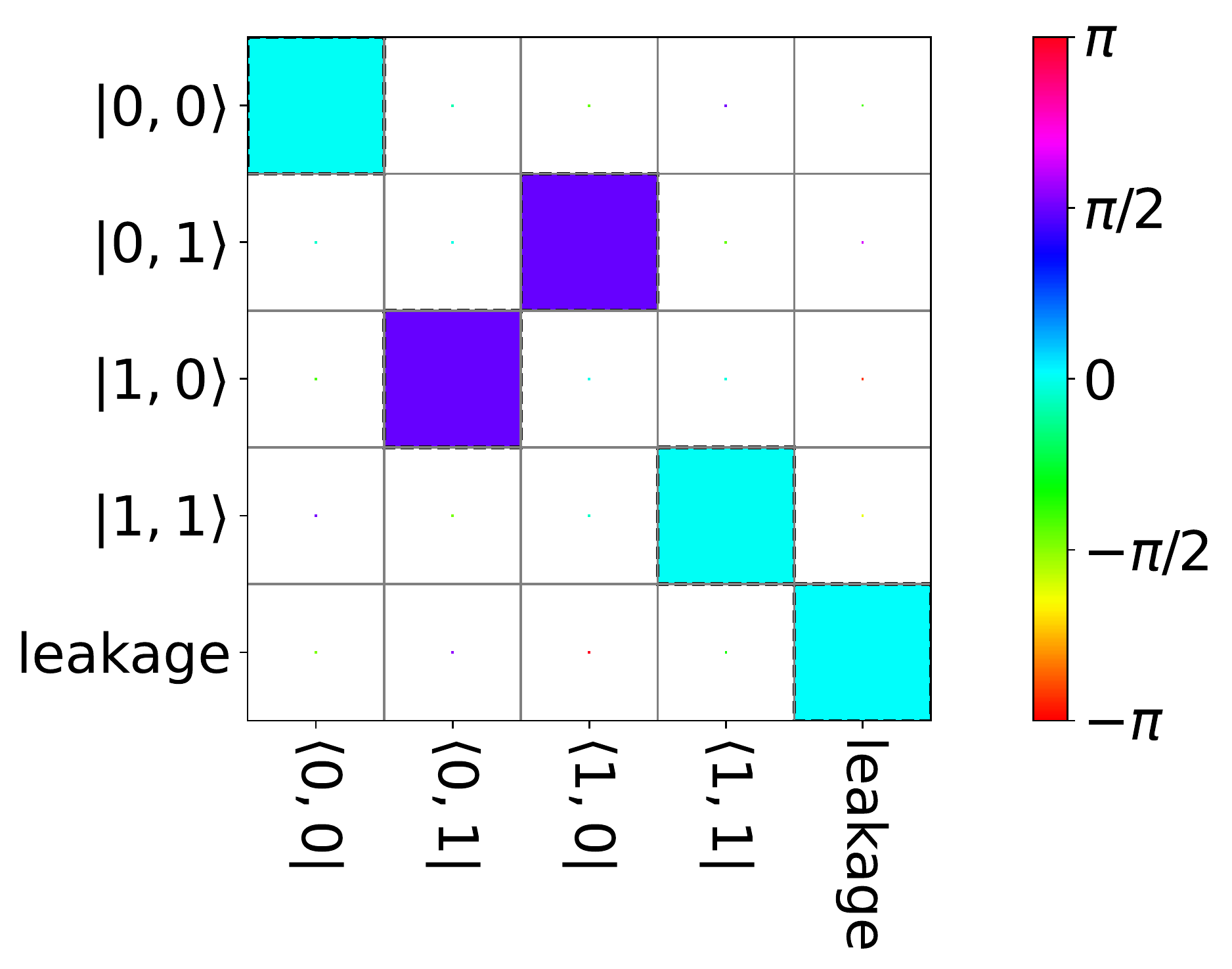}
    \end{subfigure}
    \hfill
    \begin{subfigure}[b]{0.45\textwidth}
         \centering
         \includegraphics[width=\textwidth]{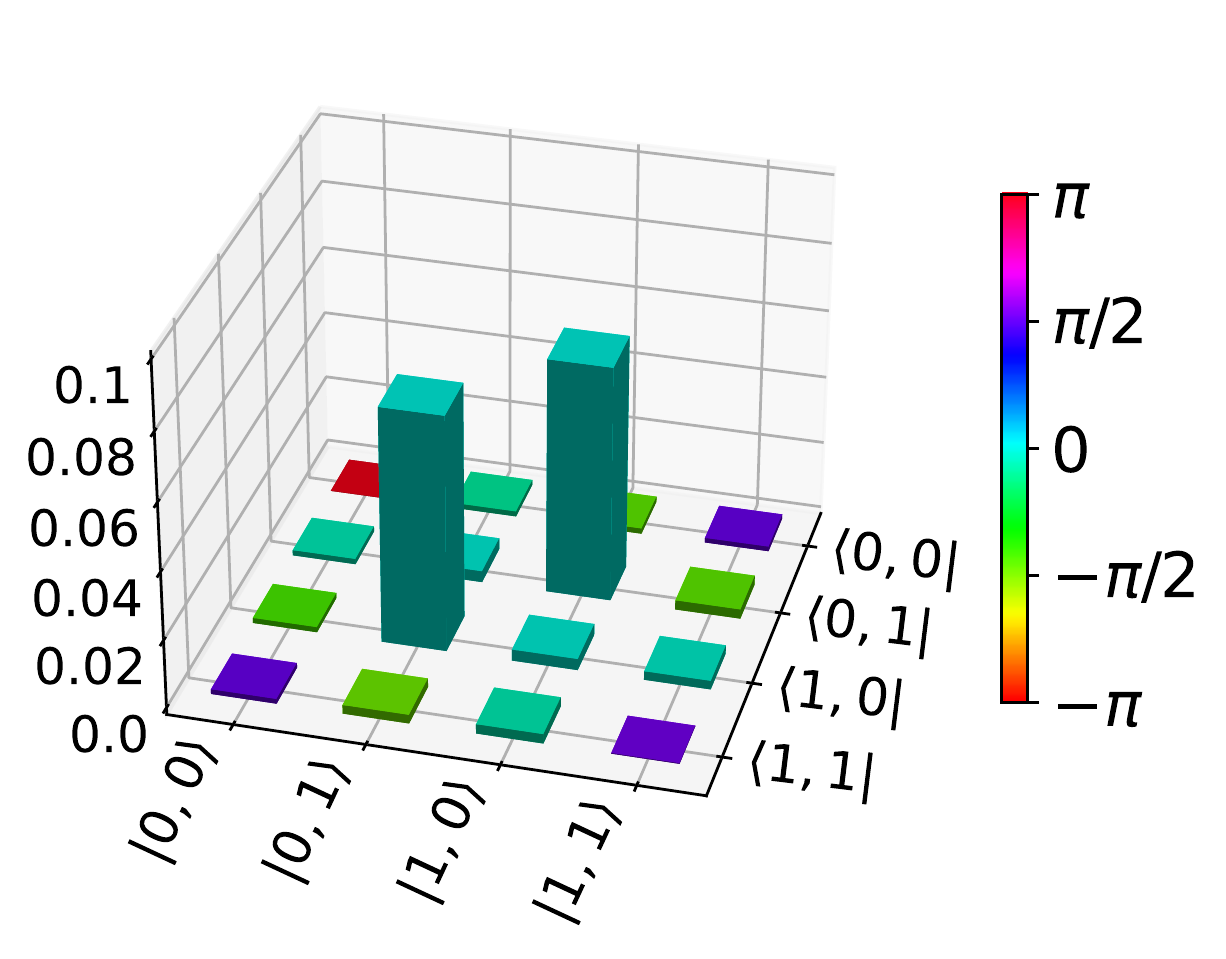}
    \end{subfigure}
    \\
    \begin{subfigure}[b]{0.48\textwidth}
        \includegraphics[width=\textwidth]{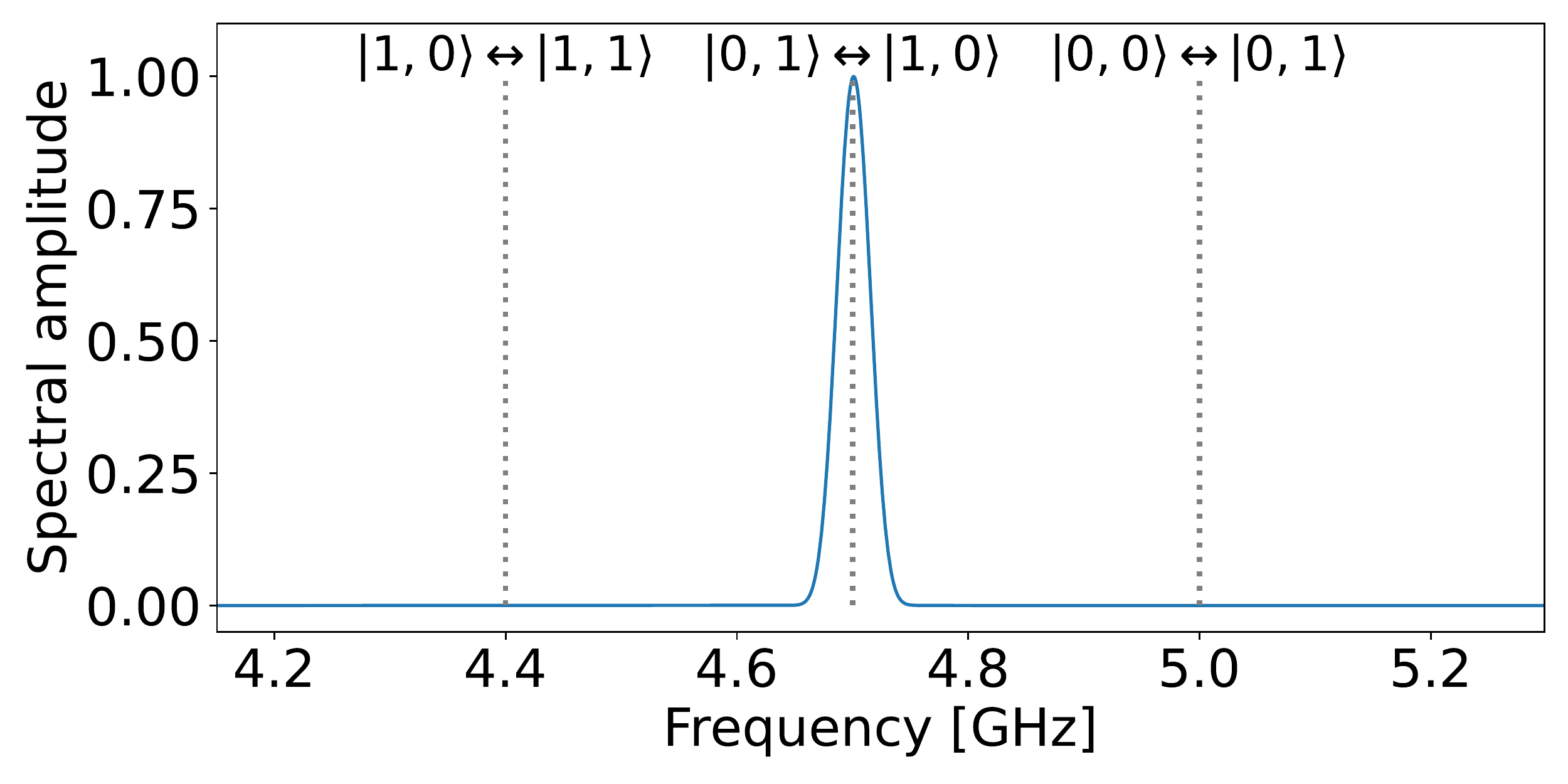}
    \end{subfigure}
    \caption{iSWAP gate between $Q_1$ and $Q_2$. Colours indicate the phase and areas are proportional to the absolute amplitude. \textbf{(a)} The optimised propagator $U$. Dotted black squares indicate the absolute values of the ideal gate. \textbf{(b)} the difference $Ue^{i\phi} - G$ between $U$ and the ideal gate $G$. The phase difference $\phi = \arctan(U_{0,0}) - \arctan(G_{0,0})$ corrects the global phase which is ignored in the optimisation. \textbf{(c)} Normalised spectral amplitude of the optimised pulse. Dotted lines correspond to resonances of the model.}
    \label{fig:single_transmon_iswap}
\end{figure*}

Entangling 2-qubit gates are typically more difficult to realise than single qubit gates, have lower fidelity, and need longer gate times. The biggest advantage of our approach is that an entangling gate between the two qubits in one transmon is easy to facilitate. As can be seen in the Hamiltonian (\ref{eq:drift_hamiltonian_paulis}), the third term is an Ising-type coupling between both qubits which corresponds to an iSWAP gate $\exp(i \frac{\pi}{4} (X\otimes X + Y\otimes Y))$. We can use this by driving the middle transition $\ket{\Bar{1}} \leftrightarrow \ket{\Bar{2}}$ to a $\frac{\pi}{2}$ rotation, which is even simpler than the single-qubit gates because it needs only one carrier frequency as long as the phases are being aligned. For this we again chose the gate time of $T=40$ ns. The final propagator is shown in figure \ref{fig:appendix_single_transmon_iswap}. With a fidelity of $1 - F < 2\cdot 10^{-3}$ it is comparable to the single-qubit gates. Although the imperfect rotation is the largest absolute error (fig. \ref{fig:appendix_single_transmon_iswap} right), a non-perfect alignment of the phase of the $\ket{11}$ state is an additional source of error and would put an upper limit on the fidelity even if the rotation was perfect. For further improvement this could be corrected by a second drive between $\ket{\Bar{2}} \leftrightarrow \ket{\Bar{3}}$ or $\ket{\Bar{3}} \leftrightarrow \ket{\Bar{4}}$ doing a full $2\pi$ rotation. Similarly, a $\sqrt{iSWAP}$ gate is possible by reducing the amplitude $A$.

In addition to the iSWAP gate, it is possible to create an entangling gate that transforms every basis state into a maximally entangled Bell pair. To realise this, a superposition of three drive signals on all neighbouring transitions $\omega_{0\rightarrow 1}$, $\omega_{1\rightarrow 2}$, and $\omega_{2\rightarrow 3}$ is necessary. The upper and lower drives at 5 and 4.4 GHz force a $\pi$-rotation, thus adding the same phase on all four levels which becomes an irrelevant global phase. At the same time the middle signal drives a $\frac{\pi}{2}$ rotation at 4.7 GHz which, on its own, would realise a $\sqrt{iSWAP}$ gate. Due to the simultaneous rotations of the other levels, however, this creates the same amplitude on the $\ket{00} \leftrightarrow \ket{11}$ matrix elements. The resulting propagator, a "double-iSWAP" gate, can be seen in figure \ref{fig:appendix_single_transmon_double_iswap}. Because of the larger drive amplitudes, the errors are considerably larger than in the iSWAP gate with $1 - F < 5 \cdot 10^{-3}$

\section{Entangling gates between two transmons}
\label{section:two_transmons}
\subsubsection*{Model Hamiltonian}
\begin{figure*}[t]
    \centering
    \begin{subfigure}[b]{0.48\textwidth}
        \centering
        \includegraphics[width=\textwidth]{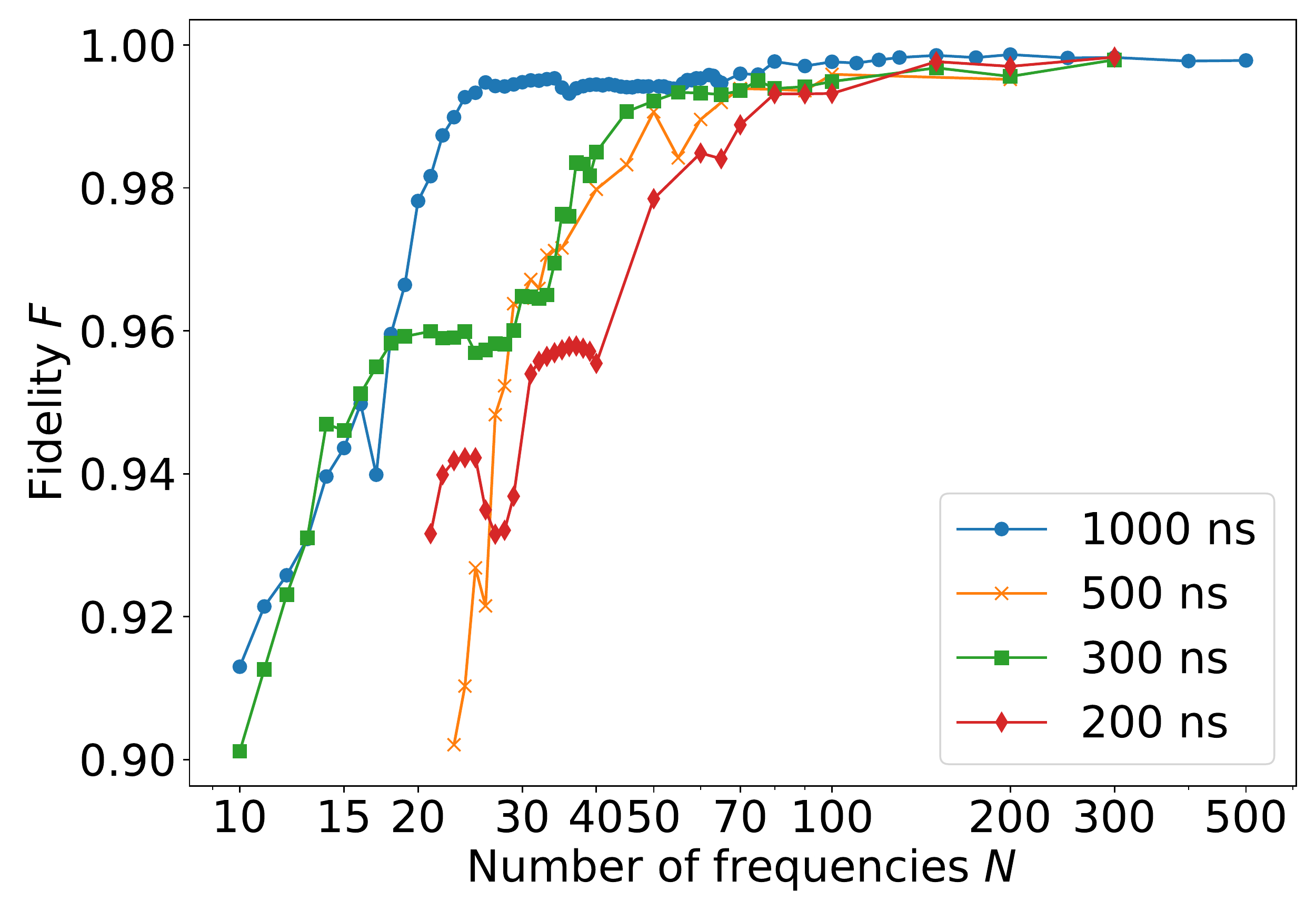}
    \end{subfigure}
    \hfill
    \begin{subfigure}[b]{0.48\textwidth}
        \centering
        \includegraphics[width=\textwidth]{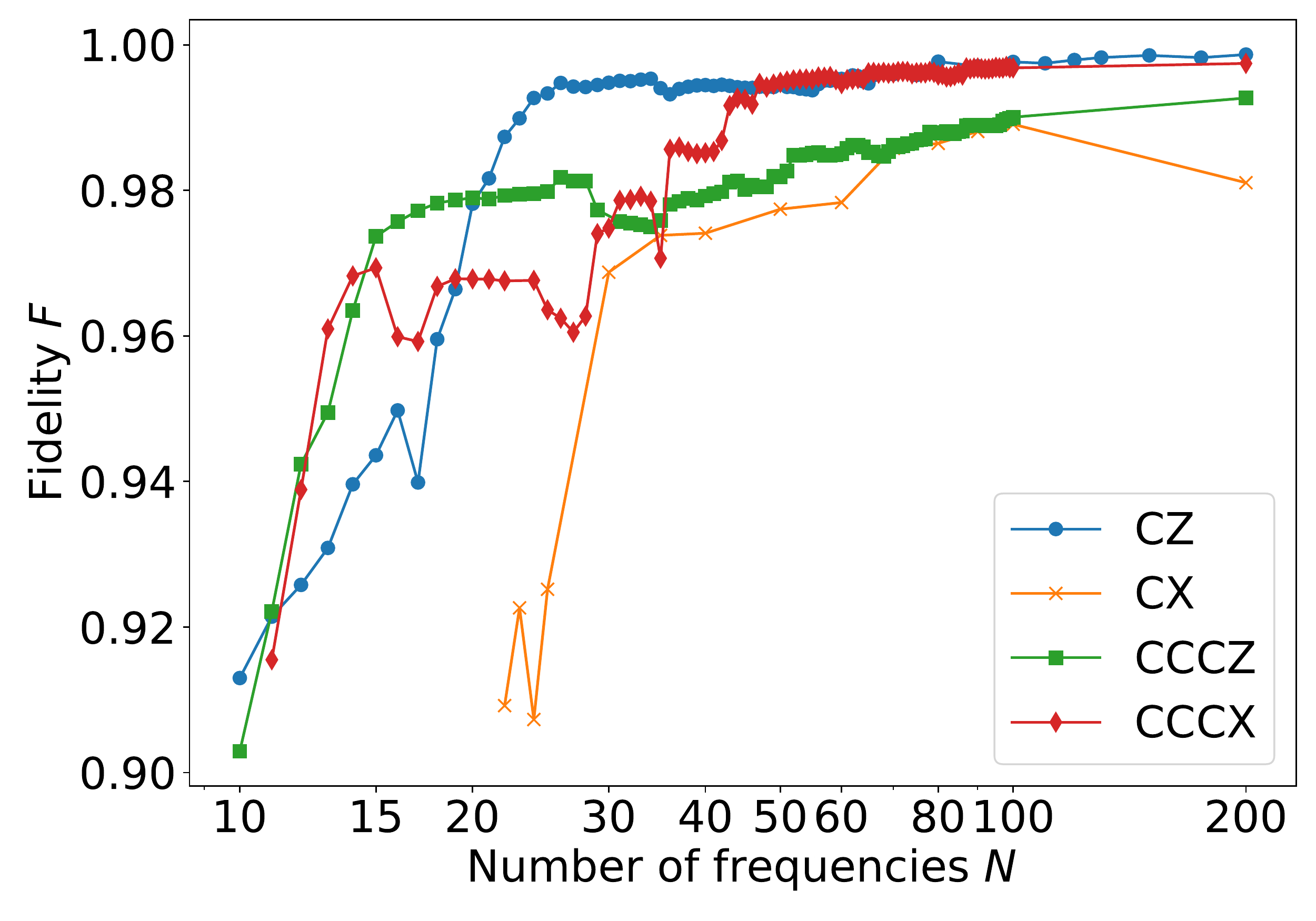}
    \end{subfigure}
    \caption{\textbf{(a)} Fidelity of the CZ gate depending on the number of frequency components. Starting from the right (200 frequencies) the number of frequencies was successively reduced and the gate was reoptimised after each step. This only includes the fidelities of $F > 0.9$. \textbf{(b)} Fidelities of 4-qubit gates CCCZ and CCCX compared to the 2-qubit gates CZ and CX.}
    \label{fig:two_transmons_fidelities_CZ}
\end{figure*}

The missing piece for a universal set of gates is at least one entangling gate between the qubits in two coupled transmons. For this we extend the model Hamiltonian to
\begin{equation}
\begin{aligned}
\label{eq:hamiltonian_two_transmons}
H &= \underbrace{\sum_{i=1,2} \left(\omega_i a_i^{\dagger} a_i + \frac{\lambda}{2} a_i^{\dagger}a_i^{\dagger}a_i a_i\right) }_{H_0} + \underbrace{\sum_{i=1,2}A_i v_i(t) (a_i+a_i^{\dagger})}_{H_d}\\ &+~ \underbrace{J (a_1 + a_1^{\dagger})(a_2 + a_2^{\dagger})}_{H_J}
\end{aligned}
\end{equation}
where the drift and drive Hamiltonian $H_0$ and $H_d$ for each of the transmons individually is the same as defined by ($\ref{eq:drift_hamiltonion_oscillator}$) and (\ref{eq:drive_hamiltonion}). In order to avoid degeneracies within the lowest 5 energy levels, the parameters for the first transmon are chosen to be $\omega_1 = 5$ GHz and $\lambda_1 = 300$ MHz like in section \ref{section:single_transmon_entangling_gate}, while the second transmon is operating at $\omega_2 = 4.5$ GHz and $\lambda_2 = 250$ MHz. We assume the transmons to be capacitively coupled with a coupling strength $J \ll \omega,\lambda$, which leads to the transverse coupling axis in (\ref{eq:hamiltonian_two_transmons}). Here, $J = 20$ MHz was used.

Due to the coupling the drift propagator $U_0$ contains many off-diagonal terms that we were not able to remove by optimisation. We therefore chose to work in the dressed basis, the eigenbasis of the coupled Hamiltonian $H_0 + H_J$, in which the propagator is diagonal but the states still accumulate phases over time. This would also be the readout basis if readout was slow. In contrast to the single transmon, it was not possible to find a gate time $T$ at which all phases align. We thus need to find a drive pulse that is able to correct all phases and, depending on the desired gate, add non-diagonal terms to the propagator.
Also, between the 25 energy levels of the combined system there are 150 unique resonances, many of which would be degenerate in the uncoupled system. The degeneracies are lifted by the coupling and are separated only by a few megahertz, turning this into a rather formidable control problem. 

\subsubsection*{Construction of pulses by reducing the number of frequencies}
To find an entangling gate with a reasonably high fidelity in this system we chose to start with a pulse shape that has sufficiently many degrees of freedom for achieving arbitrary high fidelity. After this, we successively reduced the number of freedoms and reoptimised in order to find a simpler pulse shape. First, we chose a long time of $T=1 \mu$s in order to have very narrow frequency peaks of approximately $T^{-1} = 1$ MHz which allows driving individual resonances and avoids unwanted transitions. Second, instead of two frequencies as in the single qubit case, we used a superposition of many frequencies and had the optimisation algorithm figure out which frequencies it needs. Formally this means that the drive Hamiltonian $H_d^{(i)}$ for each qubit $i$ in (\ref{eq:hamiltonian_two_transmons}) was replaced by (compare to (\ref{eq:drive_hamiltonian_superposition}))
\begin{align}
    H_d^{(i)} = &\sum_{k=1}^N A_k \cos\left(\omega_d^{(k)}t + \phi_d^{(k)}\right) (a_i + a_i^{\dagger})
\end{align}
where $i$ labels the transmon and $N$ is the number of individual frequencies. Setting $v_i(t) = 1$ corresponds to a rectangular envelope. While the cross-resonance gate was mainly facilitated by driving transmon 1 with a strong amplitude, transmon 2 was also driven with an initially weak amplitude two orders of magnitude below transmon 1 to allow for the correction of unwanted transitions. We started with $N=5000$ frequencies equally separated between 2.5 GHz and 5.5 GHz, which covers all resonances of the model, and allowed the optimiser to adjust the frequencies $\omega_d^{(k)}$, phases $\phi_d^{(k)}$, and amplitudes $A_k$. When the optimisation converged to a high fidelity, $N$ was reduced by removing frequencies with the lowest amplitudes $A_k$ (assuming that those have the least effect on the propagator) and the optimisation was started again. This successively boiled the drive Hamiltonian down to a few necessary frequencies.

\begin{figure*}[t]
    \centering
    \begin{subfigure}[b]{\textwidth}
        \includegraphics[width=\textwidth]{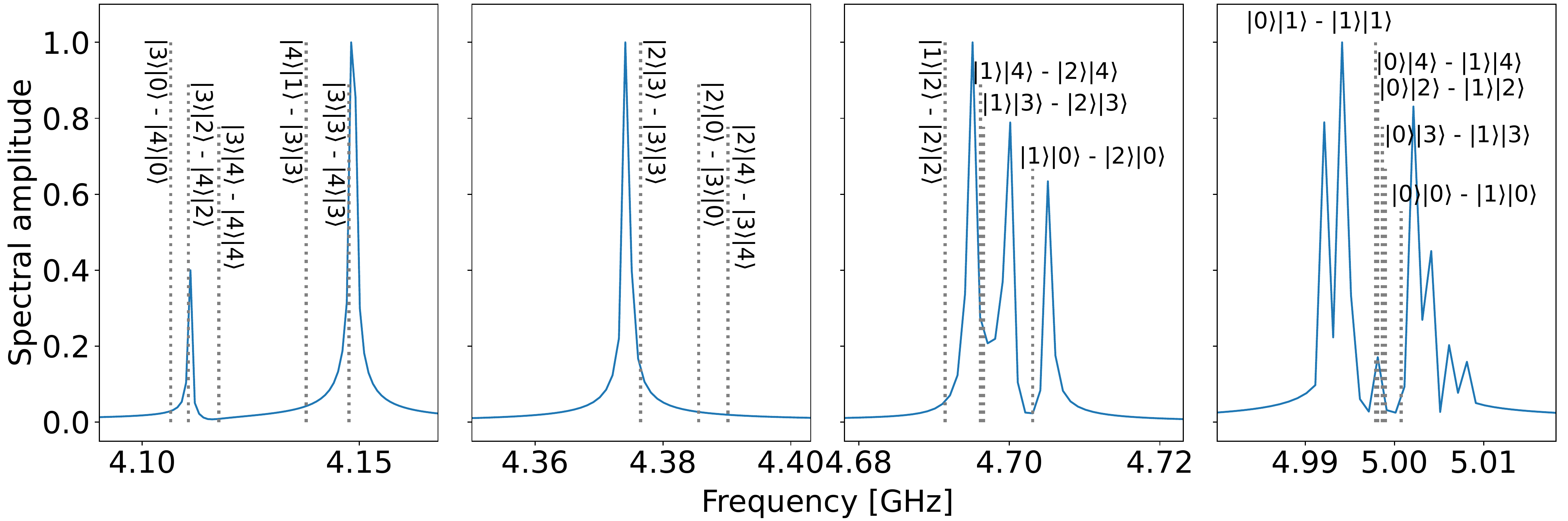}
    \end{subfigure}
    \begin{minipage}[c]{0.5\textwidth}
        \centering
        \includegraphics[width=0.9\textwidth]{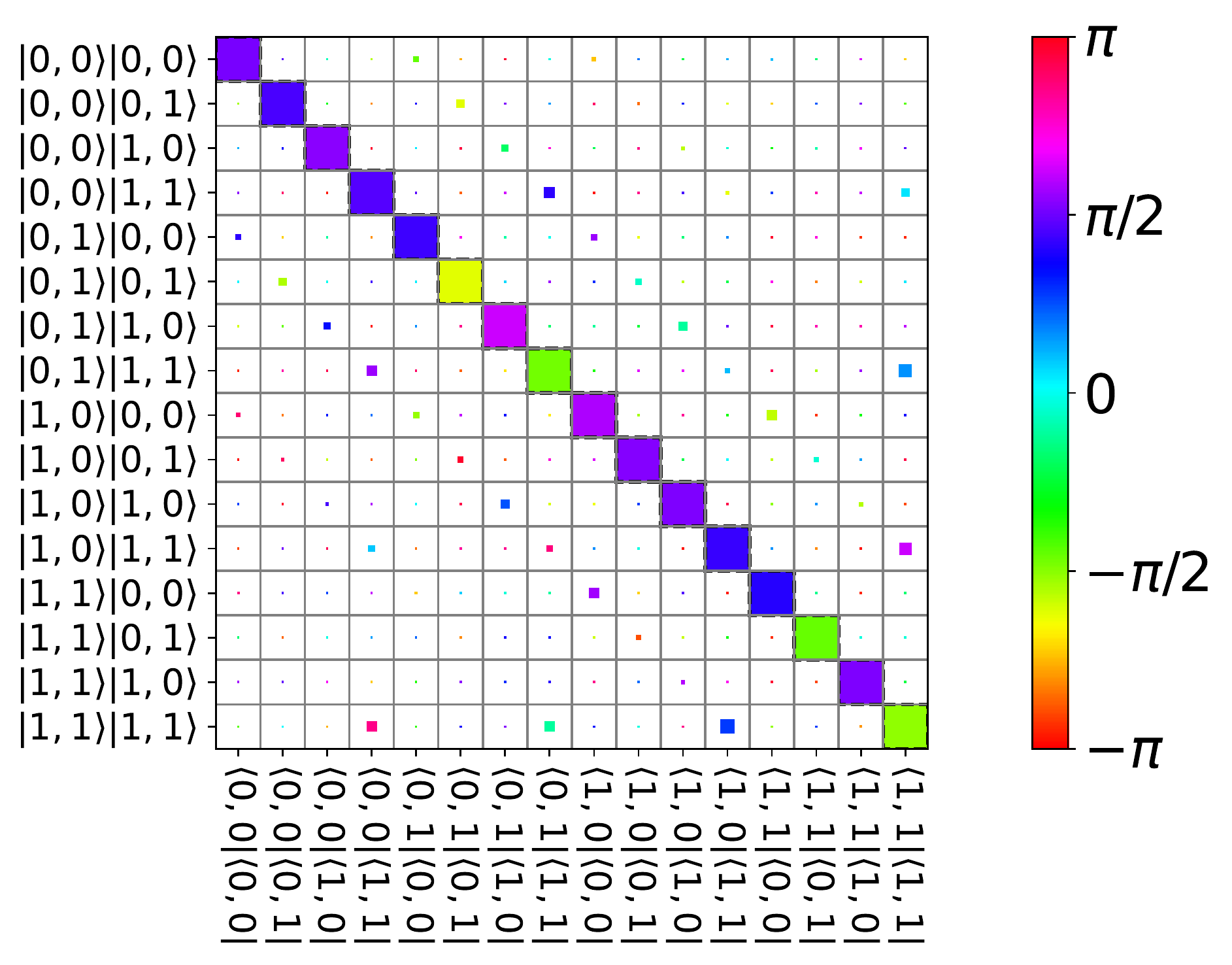}
    \end{minipage}\hfill
    \begin{minipage}[c]{0.5\textwidth}
        \caption{CZ gate between the $Q_2$ qubits of two coupled transmons that is created by $N=10$ frequencies with an infidelity of $1 - F \approx 0.035$. \textbf{(a)}: Normalised spectral amplitude of the optimised pulse on transmon 1. Dotted lines with labels correspond to Stark-shifted resonances of the model. \textbf{(b)}: The optimised propagator $U$. Dotted black squares indicate the absolute values of the ideal gate.}
        \label{fig:two_transmons_CZ_propagator}
    \end{minipage}
\end{figure*}
\subsubsection*{Realised gates}
Since the single-qubit gates on $Q_2$ were more simple than on $Q_1$, we chose to implement cross-resonance entangling gates between the $Q_2$ qubits of both transmons. In this case both $Q_1$ qubits are spectators whose state needs to stay fixed during the gate. The simplest choice for an entangling gate was the diagonal controlled-Z gate because the propagator is already diagonal in the dressed basis. For this gate, the fidelity for each number of frequency components $N$ during the reduction is shown in figure \ref{fig:two_transmons_fidelities_CZ}, where the same procedure was also done for shorter gate times of $T=200$, $300$, and $500$ ns. During the reduction the fidelity continually decreases. As can be expected, shorter gate times reach worse fidelities because the broadened spectral peaks cause unwanted transitions. Depending on the desired fidelity, 10 or less frequency components can be sufficient to realise a CZ gate. The propagator with $1 - F \approx 0.081$ that is created by 10 frequencies is depicted in Fig. \ref{fig:two_transmons_CZ_propagator}, including the model's resonances in the vicinity of the frequencies.

\subsubsection*{4-qubit gates}
In additional to 2-qubit gates with two spectators, the setup makes it possible to realise 4-qubit gates. Simple choices are triple-controlled gates in which a Pauli matrix acts on the fourth qubit only if the other three are in the state $\ket{111}$, such as controlled-Z (CCCZ) and controlled-NOT (CCCX). The propagators were calculated with the same method as above. As can be seen in  Fig. (\ref{fig:two_transmons_fidelities_CZ}), the 4-qubit gates reach similar fidelities as the 2-qubit controlled gates with the same number of frequencies.

\section{Conclusion}
\label{section:conclusion}
Using optimised pulse shapes we showed that it is possible to store and control two logical qubits within a transmon, effectively doubling the number of available qubits. The gate times of single-qubit and two-qubit gates in the transmon are comparable to usual gates in two-level systems and could be made even shorter when considering the conditions for phase alignment. Additionally, the necessary pulses with two superposed frequencies should be simple enough to be realisable experimentally. One main advantage of this approach is the fast, high-fidelity intra-transmon entangling gate. One possible caveat of our approach are shorter decoherence times of higher excited levels. Simulating the model as an open system could check if the gates are fast enough to be usable.

Entangling gates between logical qubits in two coupled transmons are more difficult to realise, need a pulse shape composed of many frequencies, and have lower fidelity. In this case, the main caveat are long gate times. Our example with a gate time of 1 $\mu s$ reaches high fidelities but is too long for real systems with noise. Although the fidelity drops with decreasing gate times, shorter gates can still be possible. It might also be possible that some values of the gate time are preferable to others. There are several ways how the gate fidelity could be improved, for example by using optimising piece-wise constant signals instead of analytical functions or by using tunable couplers instead of cross-resonance gates at fixed coupling. In general, it remains to show if the coupled system is fully controllable.

A further possibility is to assemble desired gates out of single-qubit rotations and the four-qubit entangling gates that have a comparably good fidelity even at shorter gate times. Also, using the entanglement fidelity or Makhlin invariants as the optimiser's goal function (as described in \cite{Watts_2015}) could lead to high-fidelity entangling gates other than usual choices like the CZ gate.

While working on this we became aware of two other groups who have been working on a similar idea: \cite{Fischer_2022}, \cite{Cao_2023}.

\section{Acknowledgements}
We would like to thank Nicolas Wittler and Ashutosh Mishra for fruitful discussions. This project was funded by the German Federal Ministry of Education and Research within the program  "quantum technologies - from basic research to the market" within the project GeQCos  (contract number 13N15680).

\bibliographystyle{apsrev4-1}
\bibliography{main}

\clearpage
\onecolumngrid
\appendix
\section{Optimised parameters for the gates}
\label{section:appendix_parameters}
This appendix lists the data for each optimised pulse:
\begin{itemize}
    \item A plot of the optimised propagators in which colours indicate the phase and areas are proportional to the absolute amplitude. Dotted black squares indicate the absolute values of the ideal gate.
    \item A 3D plot of the errors of the propagator with respect to the ideal gate, i.e. the difference $Ue^{i\phi} - G$ between $U$ and the ideal gate $G$. The phase difference $\phi = \arctan(U_{0,0}) - \arctan(G_{0,0})$ corrects the global phase which is ignored in the optimisation.
    \item The normalised spectral amplitude of the optimised pulses. Dotted lines show the resonances of the model.
    \item A table that lists all optimised parameter values.
\end{itemize}
The appendix contains
\begin{itemize}
    \item \ref{subsec:appendix_rx90p_q1}: $X(\pi/2)$ gate on $Q_1$
    \item \ref{subsec:appendix_rz90p_q1}: $Z(\pi/2)$ gate on $Q_1$
    \item \ref{subsec:appendix_rx90p_q2}: $X(\pi/2)$ gate on $Q_2$
    \item \ref{subsec:appendix_rz90p_q2}: $Z(\pi/2)$ gate on $Q_2$
    \item \ref{subsec:appendix_iswap}: iSWAP gate between $Q_1$ and $Q_2$
    \item \ref{subsec:appendix_double_iswap}: double-iSWAP gate between $Q_1$ and $Q_2$
    \item \ref{subsec:appendix_two_transmons}: CZ gate between both $Q_2$ qubits in two coupled transmons
\end{itemize}

% =======================================================================
\subsection{Single transmon: \texorpdfstring{$X(\pi/2)$}{} on Q1}
\label{subsec:appendix_rx90p_q1}
\begin{minipage}{0.48\textwidth}
    \centering
    \includegraphics[width=\textwidth]{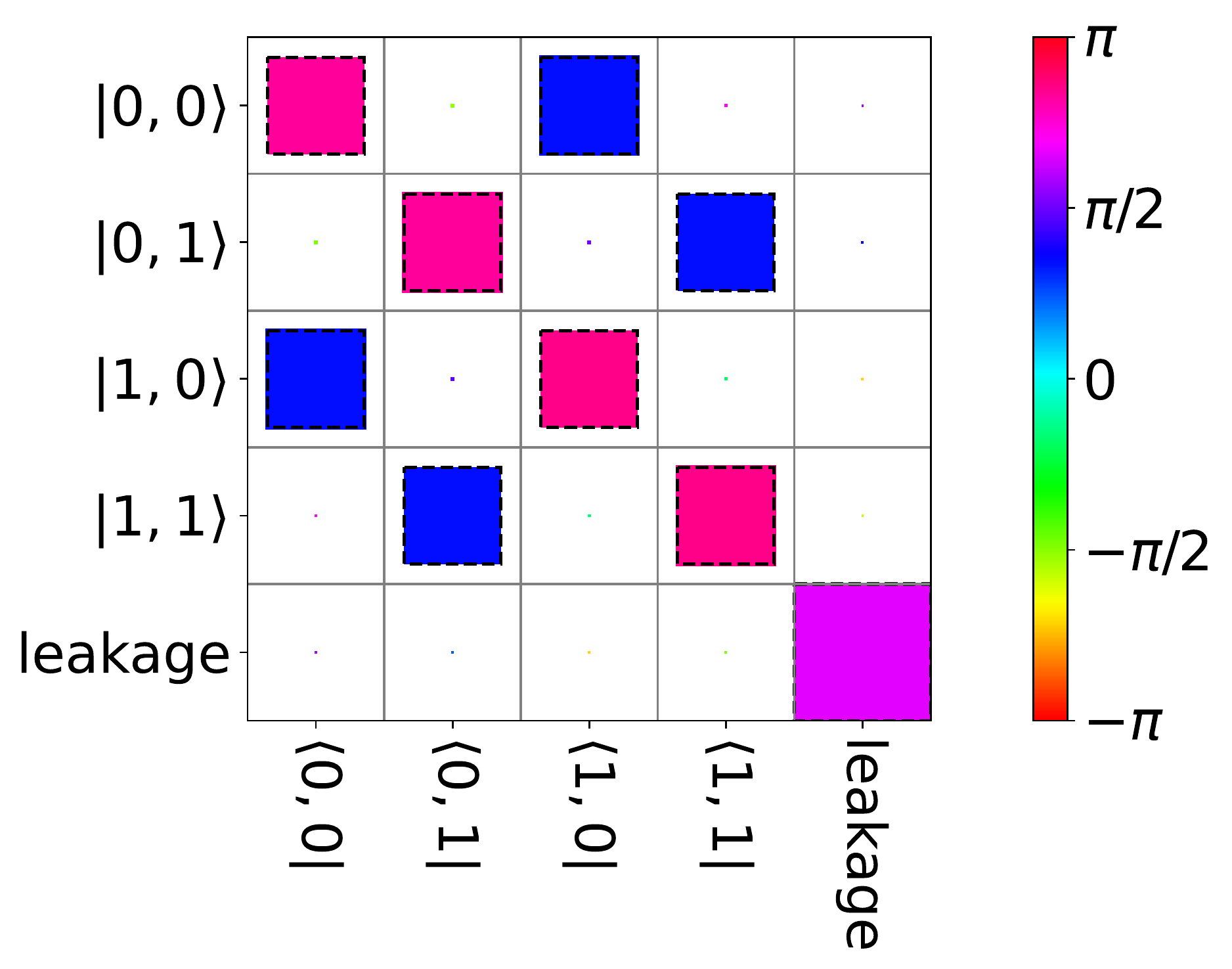}
    \label{fig:appendix_single_transmon_rx90p_q1}
\end{minipage}\hfill
\begin{minipage}{0.48\textwidth}
    \centering
    \includegraphics[width=\textwidth]{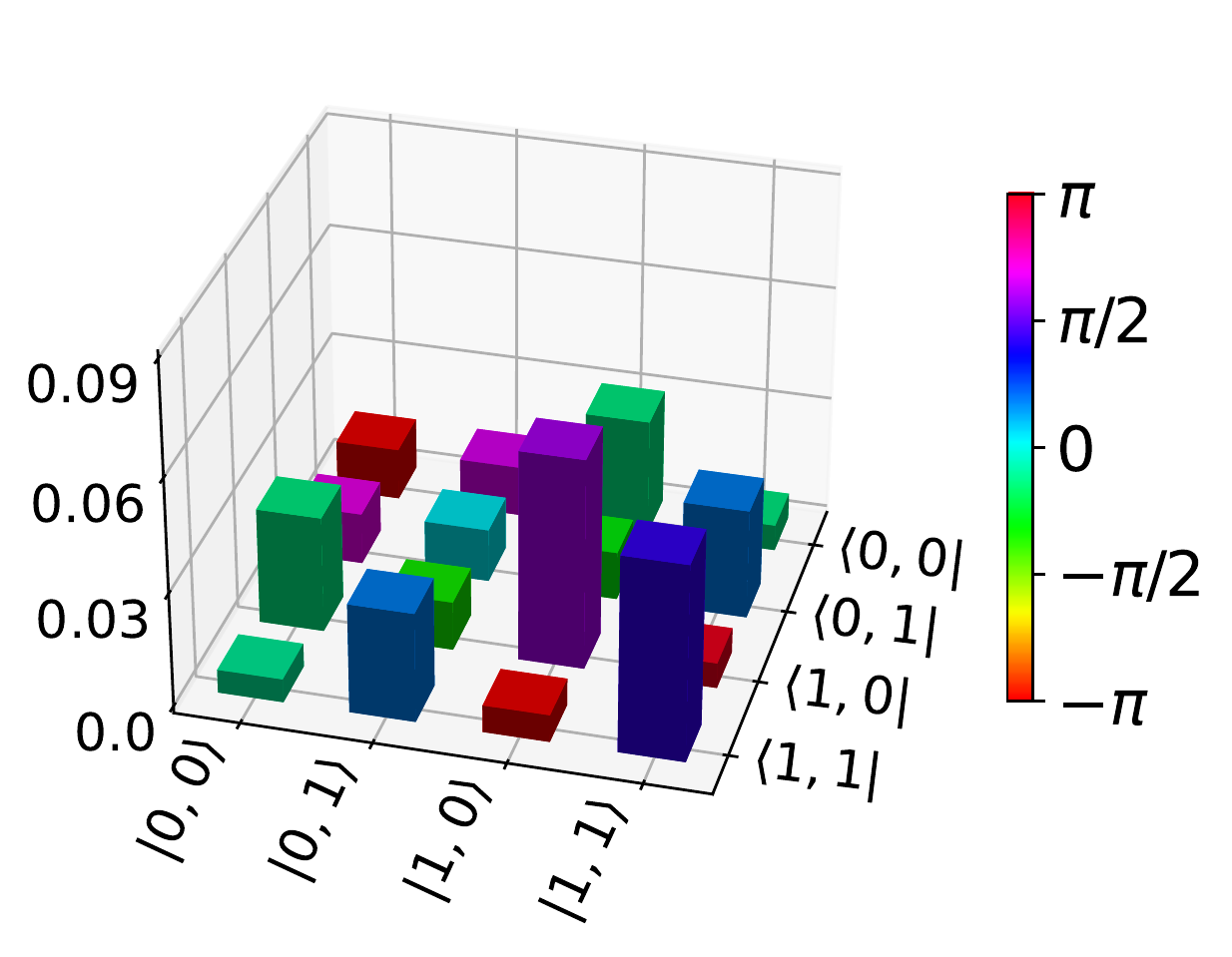}
\end{minipage}

\begin{minipage}{0.48\textwidth}
    \centering
    \includegraphics[width=\textwidth]{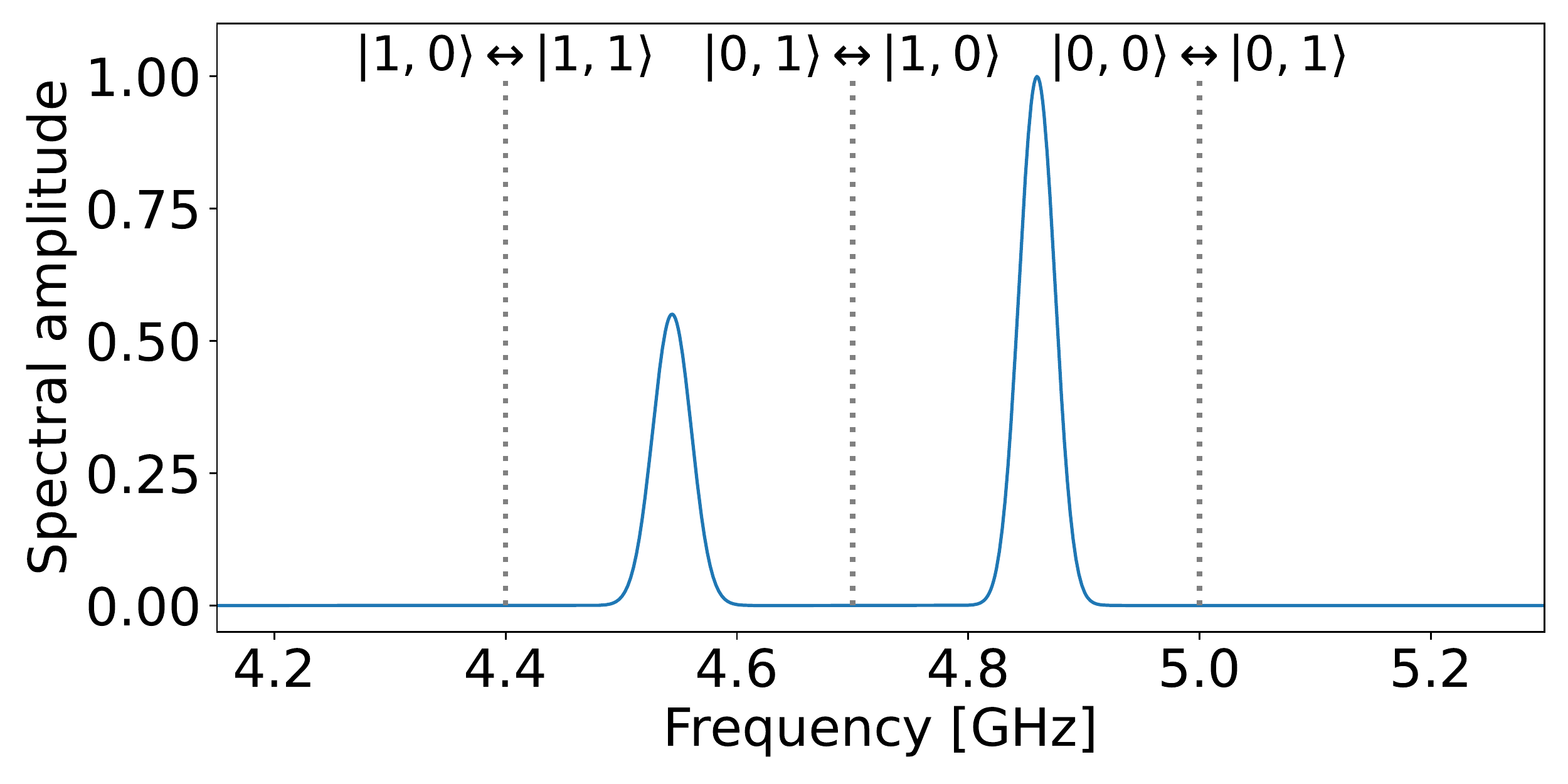}
\end{minipage}
\begin{minipage}{0.48\textwidth}
    \centering
    \begin{tabular}{c|c|c}
         ~ & Drive 1 & Drive 2 \\\hline
         Drive frequency $\omega_d$ & 4.859 GHz & 4.543 GHz \\
         Drive phase $\phi_d$ & -1.11 & 0.841 \\
         Amplitude $A$ & 638.170 mV & 511.418 mV \\
         Gaussian width $\sigma$ & 7.4 ns & 6.82 ns \\
         Gaussian center $t_0$ & 20 ns & 20 ns \\
         DRAG parameter $\delta$ & -0.0784 & 0.0432 \\
         \hline
         Gate time $T$ & \multicolumn{2}{c}{40 ns} \\
         Infidelity $1-F$ & \multicolumn{2}{c}{$1.25 \cdot 10^{-3}$}
    \end{tabular}
    \label{tab:parameters_rx90p_q1}
\end{minipage}

% =======================================================================
\subsection{Single transmon: \texorpdfstring{$Z(\pi/2)$}{} on Q1}
\label{subsec:appendix_rz90p_q1}
\begin{minipage}{0.48\textwidth}
    \centering
    \includegraphics[width=\textwidth]{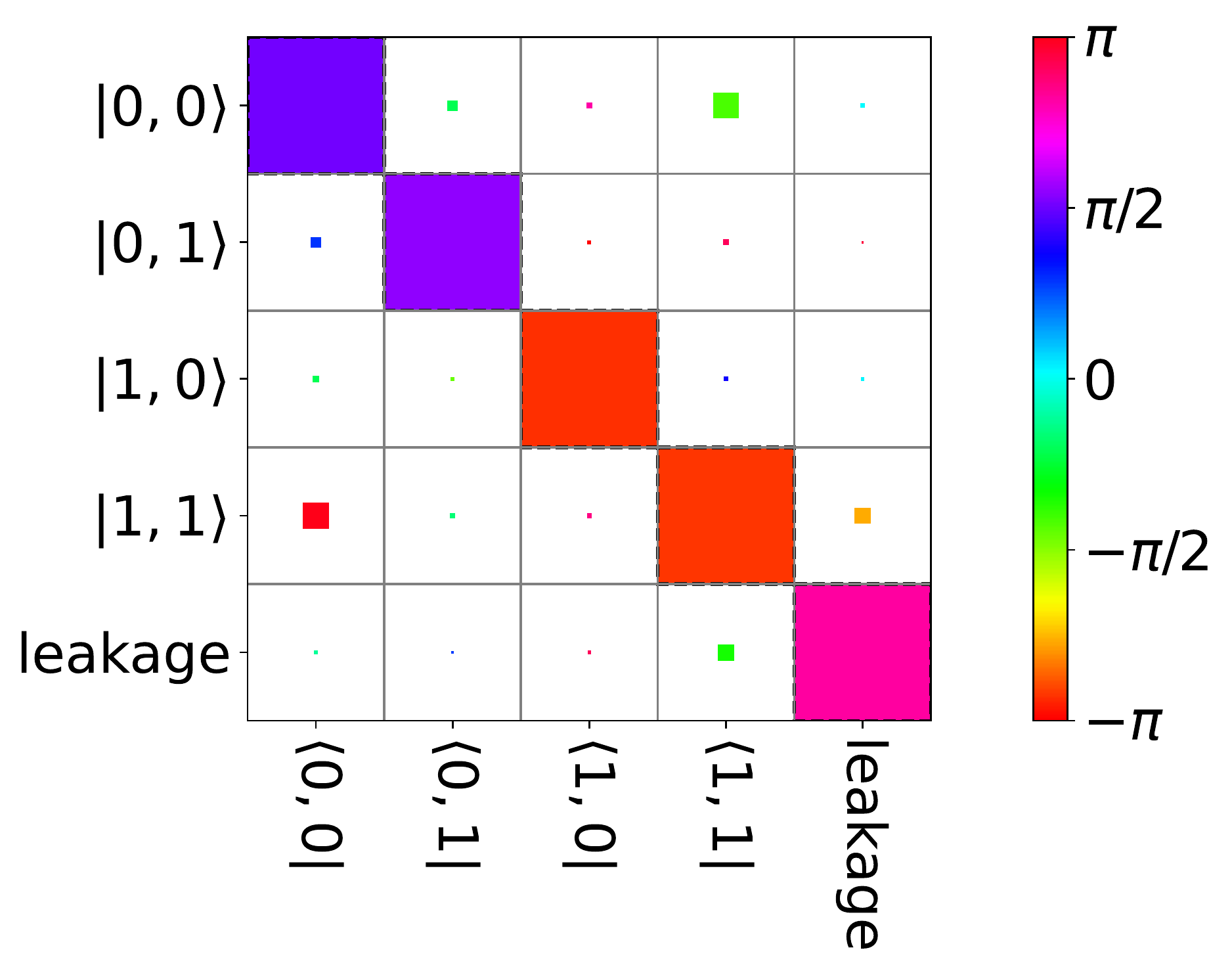}
    \label{fig:appendix_single_transmon_rz90p_q1}
\end{minipage}\hfill
\begin{minipage}{0.48\textwidth}
     \centering
     \includegraphics[width=\textwidth]{img_single_transmon/rz90p_q1/rz90p_q1_propagator_diff.pdf}
\end{minipage}

\begin{minipage}{0.48\textwidth}
    \centering
    \includegraphics[width=\textwidth]{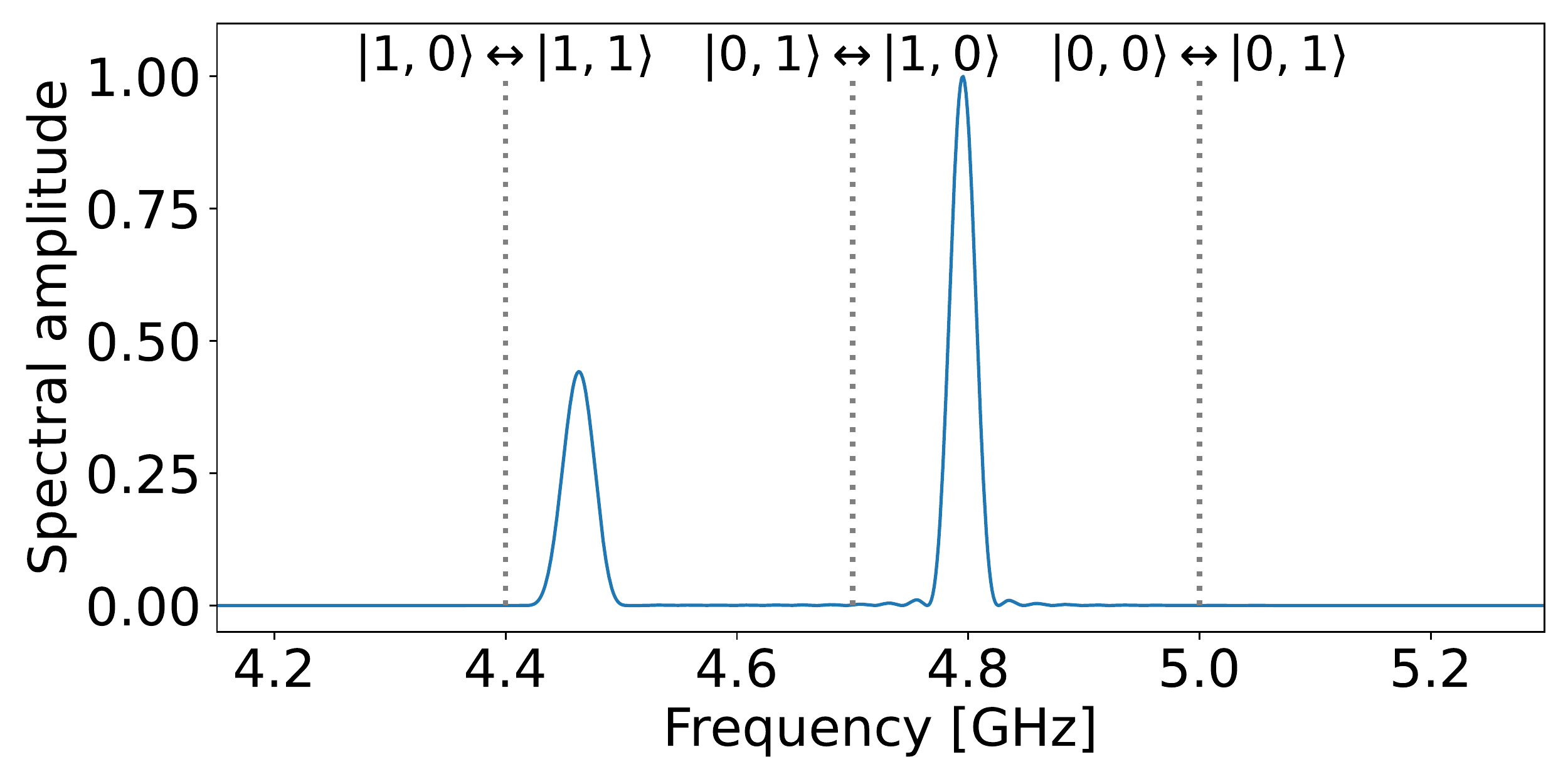}
\end{minipage}\hfill
\begin{minipage}{0.48\textwidth}
    \centering
    \begin{tabular}{c|c|c}
         ~ & Drive 1 & Drive 2 \\\hline
         Drive frequency $\omega_d$ & 4.795 GHz & 4.462 GHz \\
         Drive phase $\phi_d$ & 1.418 & -0.8141 \\
         Amplitude $A$ & 492.850 mV & 446.878 mV \\
         Gaussian width $\sigma$ & 14.660 ns & 9.284 ns \\
         Gaussian center $t_0$ & 19.8985 ns & 19.956 ns \\
         DRAG parameter $\delta$ & 1.0 & 1.0 \\
         \hline
         Gate time $T$ & \multicolumn{2}{c}{40 ns} \\
         Infidelity $1-F$ & \multicolumn{2}{c}{$0.026$}
    \end{tabular}
    \label{tab:parameters_rz90p_q1}
\end{minipage}

% =======================================================================
\subsection{Single transmon: \texorpdfstring{$X(\pi/2)$}{} on Q2}
\label{subsec:appendix_rx90p_q2}
\begin{minipage}{0.48\textwidth}
    \centering
    \includegraphics[width=\textwidth]{img_single_transmon/rx90p_q2/rx90p_q2_propagator.pdf}
    \label{fig:appendix_single_transmon_rx90p_q2}
\end{minipage}\hfill
\begin{minipage}{0.48\textwidth}
     \centering
     \includegraphics[width=\textwidth]{img_single_transmon/rx90p_q2/rx90p_q2_propagator_diff.pdf}
\end{minipage}
\begin{minipage}{0.48\textwidth}
    \centering
    \includegraphics[width=\textwidth]{img_single_transmon/rx90p_q2/rx90p_q2_spectrum.pdf}
\end{minipage}\hfill
\begin{minipage}{0.48\textwidth}
    \centering
    \begin{tabular}{c|c|c}
         ~ & Drive 1 & Drive 2 \\\hline
         Drive frequency $\omega_d$ & 5.001 GHz & 4.401 GHz \\
         Drive phase $\phi_d$ & -0.115 & -0.071 \\
         Amplitude $A$ & 102.863 mV & 56.413 mV \\
         Gaussian width $\sigma$ & 6.124 ns & 6.451 ns \\
         Gaussian center $t_0$ & 20 ns & 20 ns \\
         DRAG parameter $\delta$ & 0 & 0 \\
         \hline
         Gate time $T$ & \multicolumn{2}{c}{40 ns} \\
         Infidelity $1-F$ & \multicolumn{2}{c}{$5.96 \cdot 10^{-7}$}
    \end{tabular}
    \label{tab:parameters_rx90p_q2}
\end{minipage}

% =======================================================================
\subsection{Single transmon: \texorpdfstring{$Z(\pi/2)$}{} on Q2}
\label{subsec:appendix_rz90p_q2}
\begin{minipage}{0.48\textwidth}
    \centering
    \includegraphics[width=\textwidth]{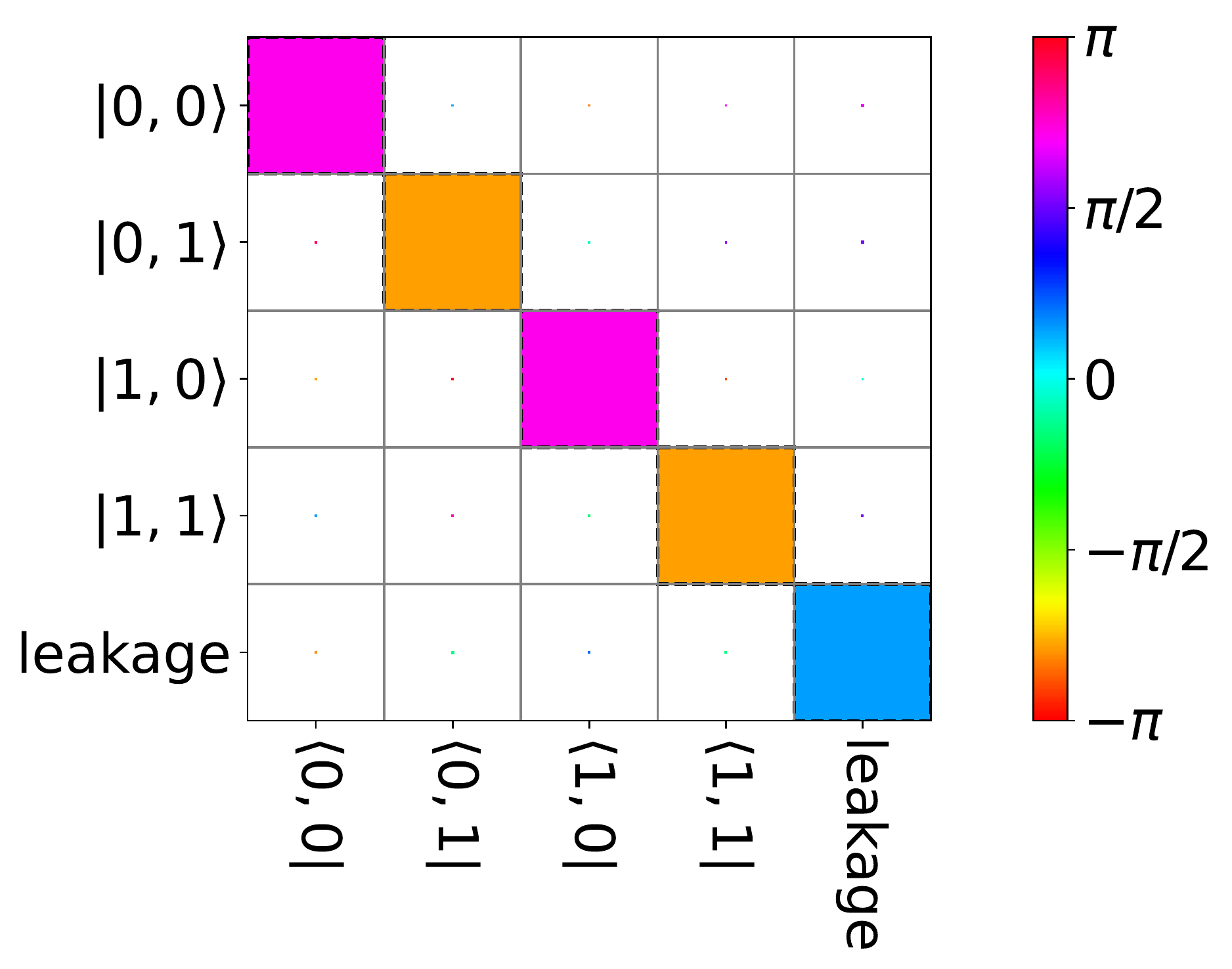}
    \label{fig:appendix_single_transmon_rz90p_q2}
\end{minipage}\hfill
\begin{minipage}{0.48\textwidth}
     \centering
     \includegraphics[width=\textwidth]{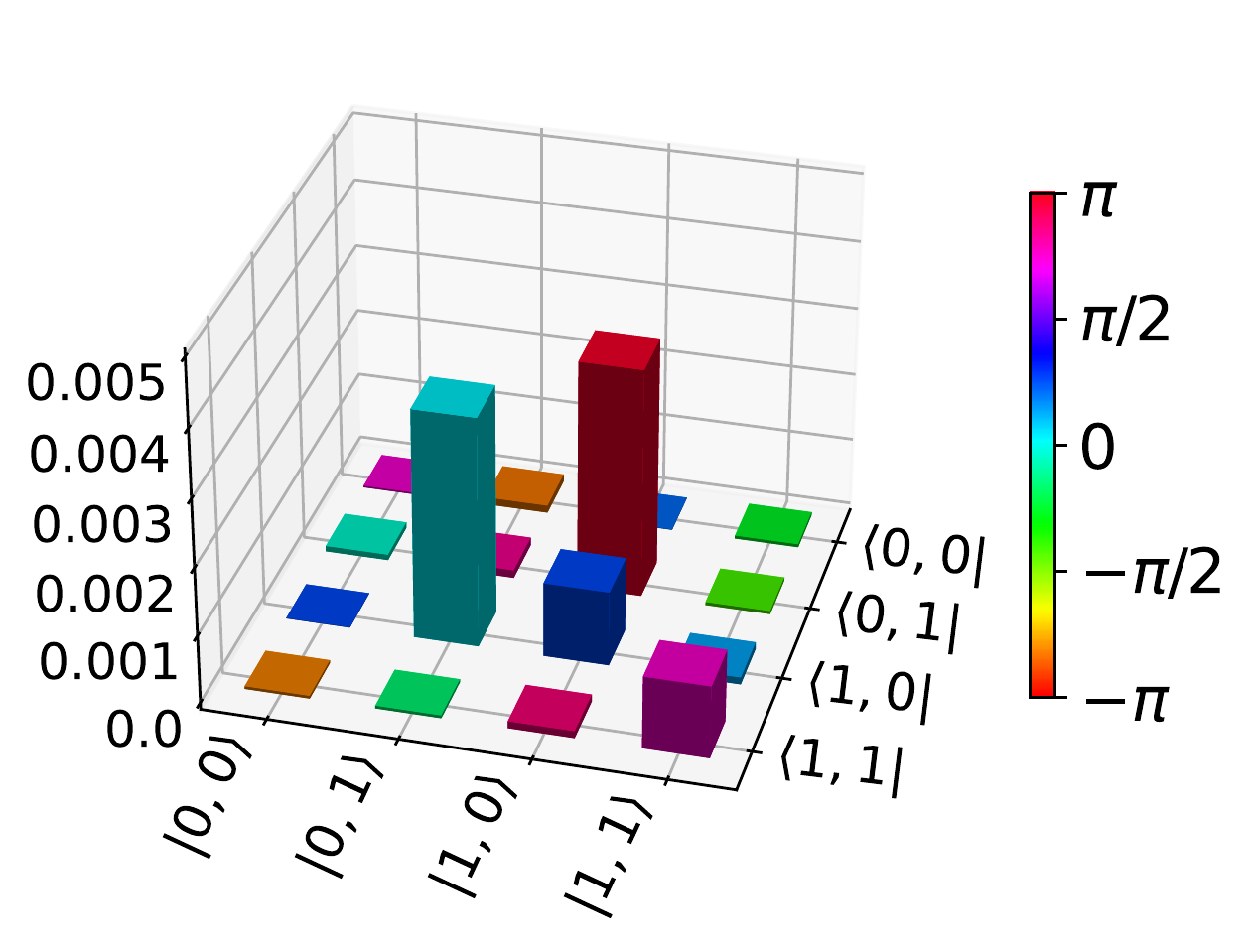}
\end{minipage}

\begin{minipage}{0.48\textwidth}
    \centering
    \includegraphics[width=\textwidth]{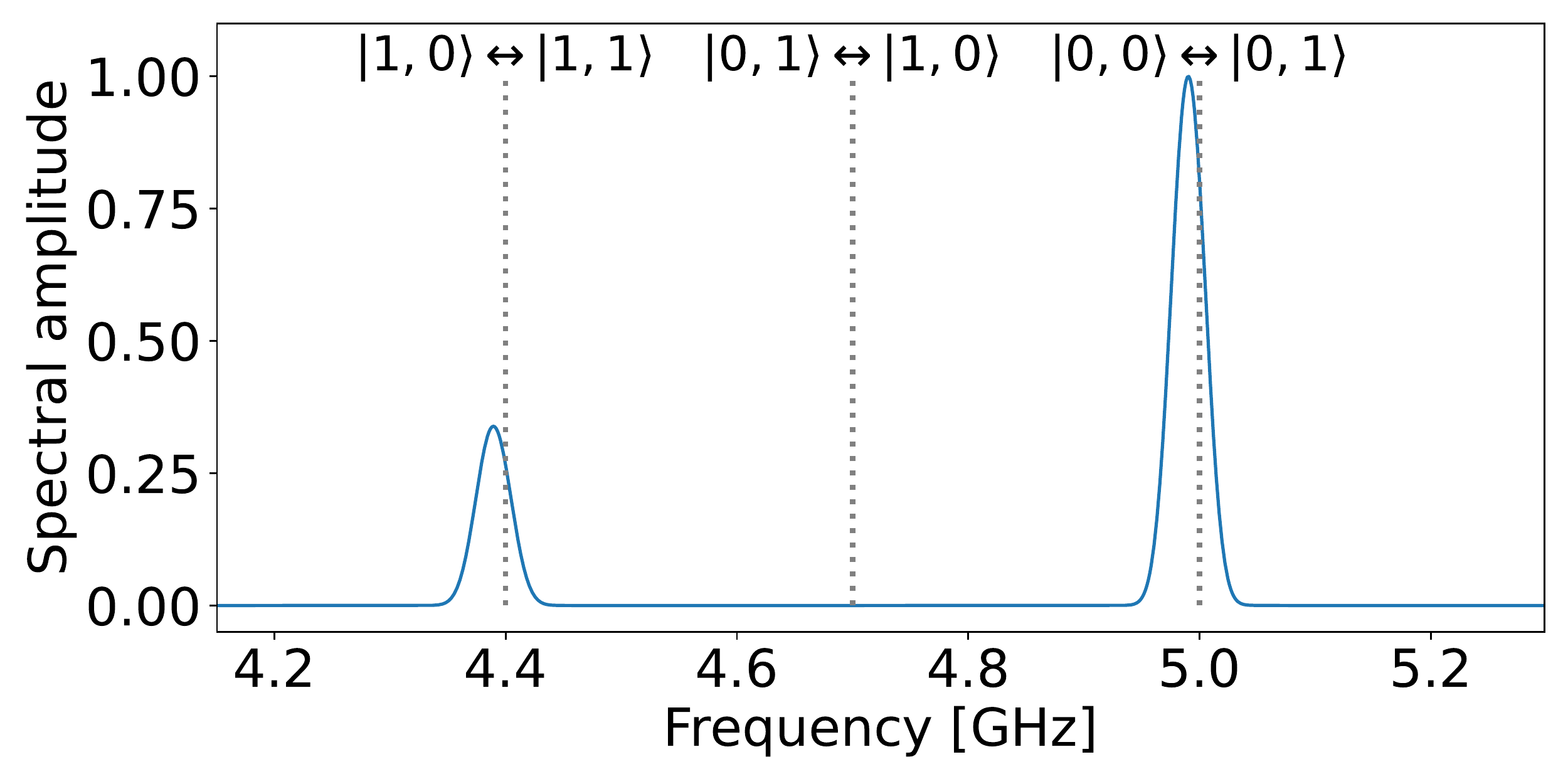}
\end{minipage}\hfill
\begin{minipage}{0.48\textwidth}
    \centering
    \begin{tabular}{c|r|r}
         ~ & Drive 1 & Drive 2 \\\hline
         Drive frequency $\omega_d$ & 4.99 GHz & 4.389 GHz \\
         Drive phase $\phi_d$ & 6.123 $\cdot 10^{-3}$ & -3.130 $\cdot 10^{-3}$\\
         Amplitude $A$ & 305.339 mV & 184.267 mV \\
         Gaussian width $\sigma$ & 8.045 ns & 7.735 ns \\
         Gaussian center $t_0$ & 20 ns & 20 ns \\
         DRAG parameter $\delta$ & -0.009354 & 0.010245 \\
         \hline
         Gate time $T$ & \multicolumn{2}{c}{40 ns} \\
         Infidelity $1-F$ & \multicolumn{2}{c}{$2.91 \cdot 10^{-5}$}
    \end{tabular}
    \label{tab:parameters_rz90p_q2}
\end{minipage}

% =======================================================================
\subsection{Single transmon: iSWAP}
\label{subsec:appendix_iswap}
\begin{minipage}{0.48\textwidth}
    \centering
    \includegraphics[width=\textwidth]{img_single_transmon/iswap/iswap_propagator.pdf}
    \label{fig:appendix_single_transmon_iswap}
\end{minipage}\hfill
\begin{minipage}{0.48\textwidth}
     \centering
     \includegraphics[width=\textwidth]{img_single_transmon/iswap/iswap_propagator_diff.pdf}
\end{minipage}

\begin{minipage}{0.48\textwidth}
    \centering
    \includegraphics[width=\textwidth]{img_single_transmon/iswap/iswap_spectrum.pdf}
\end{minipage}\hfill
\begin{minipage}{0.48\textwidth}
    \centering
    \begin{tabular}{c|c}
         ~ & Drive 1 \\\hline
         Drive frequency $\omega_d$ & 4.7 GHz \\
         Drive phase $\phi_d$ & 3.1 \\
         Amplitude $A$ & 109.661 mV \\
         Gaussian width $\sigma$ & 8.237 ns \\
         Gaussian center $t_0$ & 20 ns \\
         DRAG parameter $\delta$ & 1.048 \\
         \hline
         Gate time $T$ & 40 ns \\
         Infidelity $1-F$ & $1.242 \cdot 10^{-3}$
    \end{tabular}
    \label{tab:parameters_iswap}
\end{minipage}

% =======================================================================
\subsection{Single transmon: double iSWAP}
\label{subsec:appendix_double_iswap}

\begin{minipage}{0.48\textwidth}
    \centering
    \includegraphics[width=\textwidth]{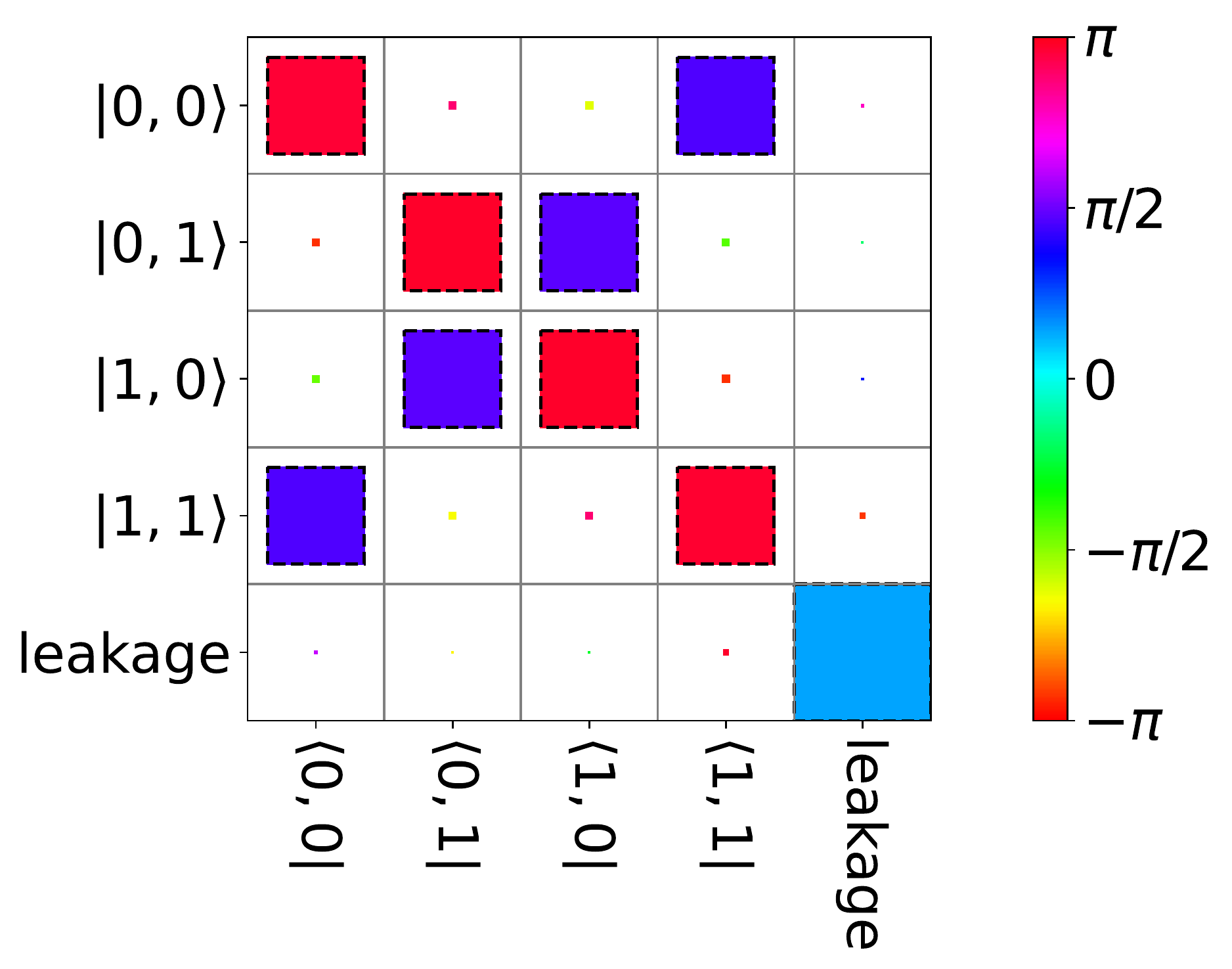}
    \label{fig:appendix_single_transmon_double_iswap}
\end{minipage}\hfill
\begin{minipage}{0.48\textwidth}
     \centering
     \includegraphics[width=\textwidth]{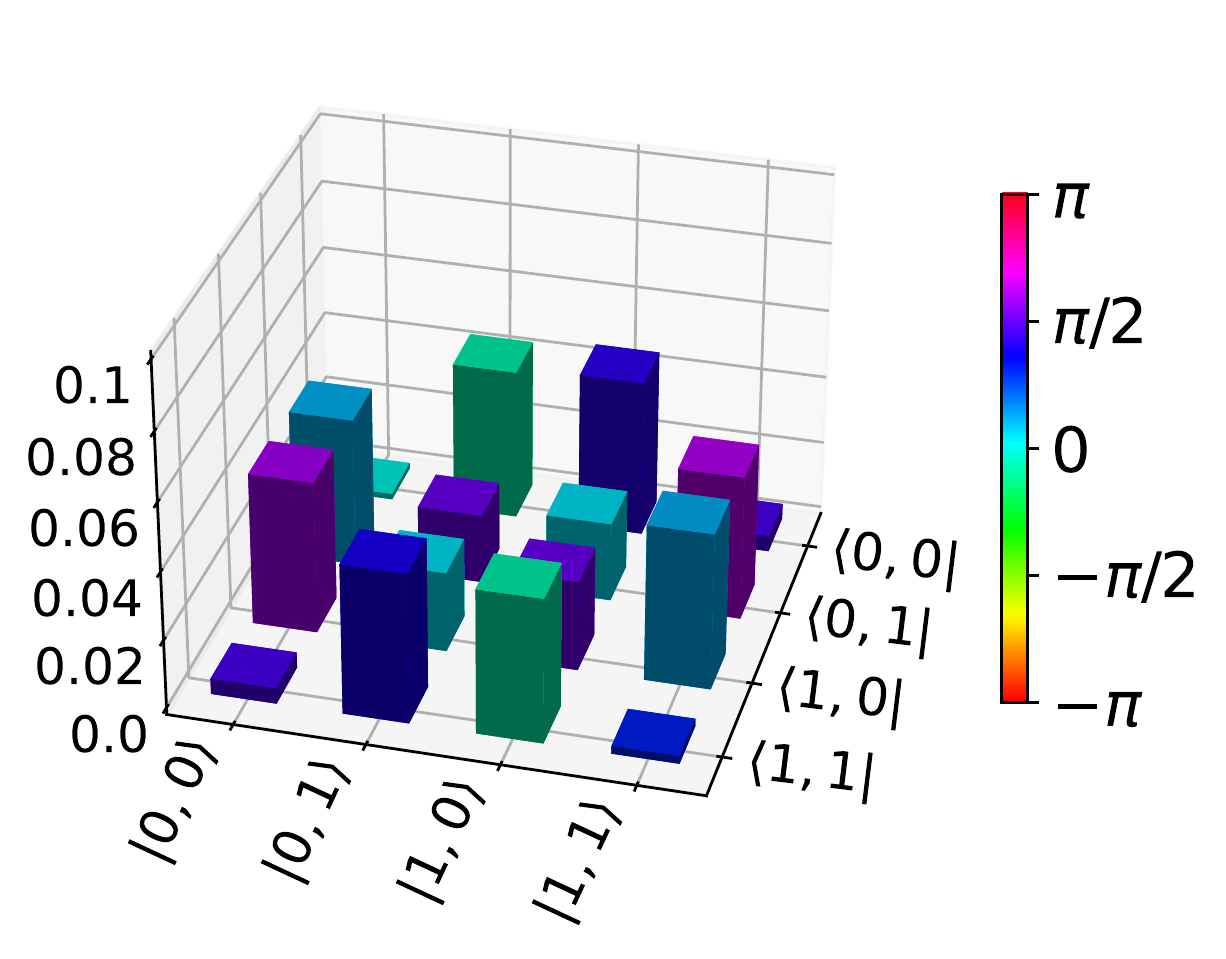}
\end{minipage}

\begin{minipage}{0.48\textwidth}
    \centering
    \includegraphics[width=\textwidth]{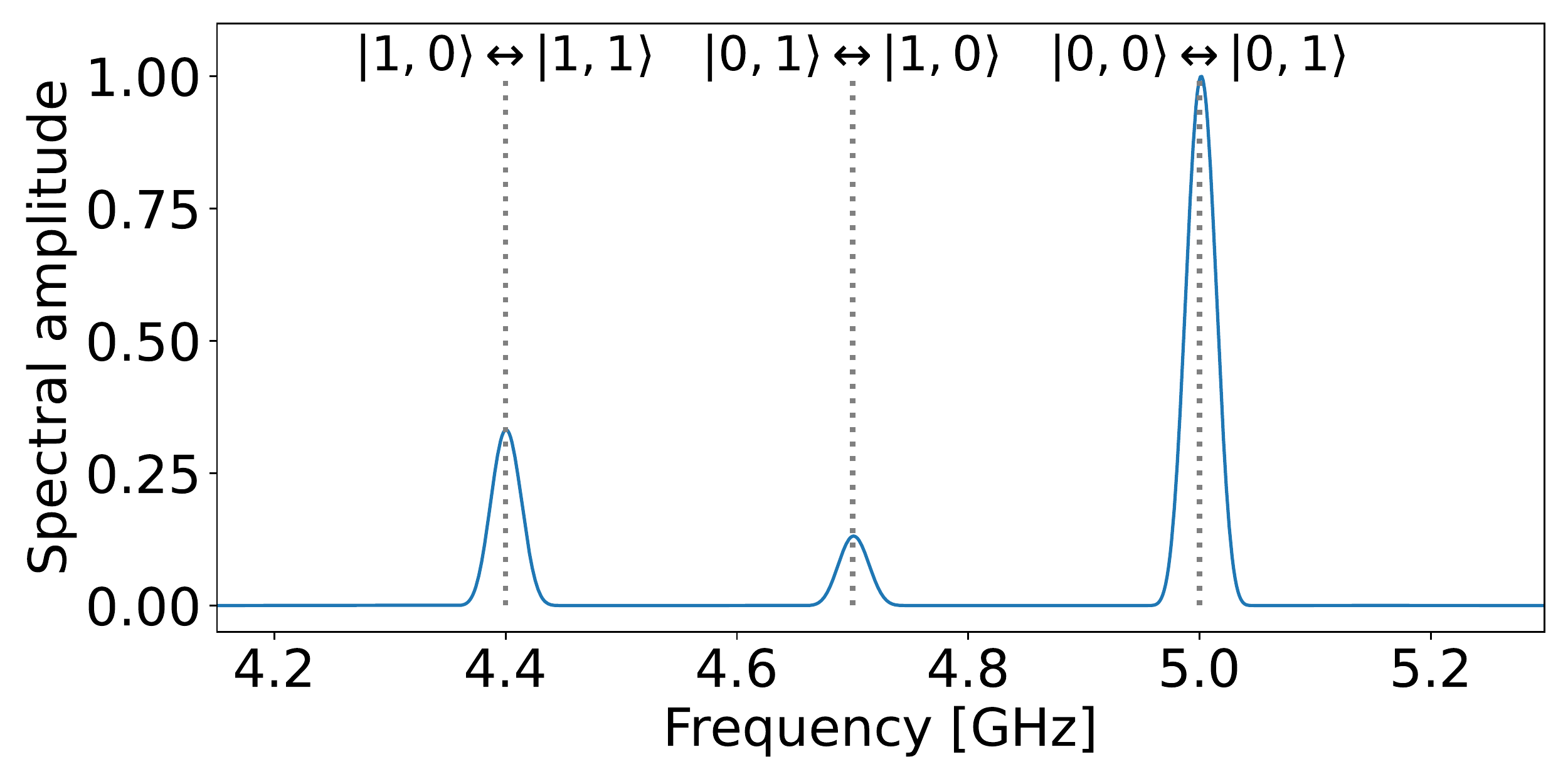}
\end{minipage}\hfill
\begin{minipage}{0.48\textwidth}
    \centering
    \begin{tabular}{c|c|c|c}
         ~ & Drive 1& Drive 2 & Drive 3 \\\hline
         Drive frequency $\omega_d$ & 5 GHz & 4.7 GHz & 4.4 GHz \\
         Drive phase $\phi_d$ & -0.0224 & -0.0651 & -0.0319 \\
         Amplitude $A$ & 264.977 mV & 96.237 mV & 154.838 mV \\
         Gaussian width $\sigma$ & 9.6 ns & 9.596 ns & 9.436 ns \\
         Gaussian center $t_0$ & 20 ns & 20 ns & 20 ns\\
         DRAG parameter $\delta$ & 0.0141 & -0.0073 & 0.0139 \\
         \hline
         Gate time $T$ & 40 ns \\
         Infidelity $1-F$ & $4.41 \cdot 10^{-3}$
    \end{tabular}
    \label{tab:parameters_double_iswap}
\end{minipage}

% =======================================================================
\subsection{Two transmons: Controlled-Z gate}
\label{subsec:appendix_two_transmons}
This is the data for the CZ gate generated by 10 frequencies at a gate time of 1000 ns.
The propagator and spectrum of the CZ gate are included in the main text in figure \ref{fig:two_transmons_CZ_propagator}.
\vspace{20px}

\begin{minipage}{0.48\textwidth}
    \centering
    \begin{tabular}{c|c|c|c|c|c|c}
        ~ & \multicolumn{3}{c|}{Transmon 1} & \multicolumn{3}{c}{Transmon 2} \\
        ~ & \makecell{Drive frequency\\ $\omega_d$ [GHz]} & \makecell{Drive phase\\$\phi_d$} & Amplitude $A$ [mV] & \makecell{Drive frequency\\$\omega_d$ [GHz]} & \makecell{Drive phase\\$\phi_d$} & Amplitude $A$ [µV] \\ 
        \hline
        Drive 1 & 4.994 & -0.982 & 50.116 & 4.106 & 0.804 & 42.356\\
        Drive 2 & 5.002 & -0.406 & 42.486 & 4.109 & -2.520 & 40.886\\
        Drive 3 & 4.374 & -2.276 & 30.302 & 4.254 & -1.438 & 40.552\\
        Drive 4 & 4.992 & -0.596 & 30.368 & 4.260 & -1.646 & 40.510\\
        Drive 5 & 4.148 & -1.120 & 23.789 & 4.260 & 1.247 & 40.512\\
        Drive 6 & 5.004 & -3.054 & 16.527 & 4.384 & -0.912 & 40.427\\
        Drive 7 & 4.695 & 1.390 & 15.052 & 4.114 & 2.275 & 40.316\\
        Drive 8 & 4.700 & -1.465 & 13.235 & 4.504 & 0.712 & 40.159\\
        Drive 9 & 4.705 & 0.697 & 9.623 & 4.382 & 0.632 & 40.153\\
        Drive 10 & 4.111 & -1.876 & 7.099 & 4.126 & -2.947 & 40.079\\
        \hline
        Gate time $T$ & 1000 ns \\
        Infidelity $1-F$ & $8.701 \cdot 10^{-2}$
    \end{tabular}
    \label{tab:parameters_two_transmons_cx}
\end{minipage}

\end{document}